\documentclass[a4paper,fleqn,usenatbib]{mnras}
\usepackage{newtxtext,newtxmath}
\usepackage{siunitx}
\usepackage[T1]{fontenc}
\usepackage{ae,aecompl}
\usepackage{graphicx}	
\usepackage{amsmath}	
\usepackage{amssymb}	
\usepackage{lscape}
\graphicspath{ {figures/} }
\usepackage{siunitx}
\usepackage{longtable}
\usepackage{hyperref}
\hypersetup{colorlinks}
\let\oldhref\href
\renewcommand{\href}[2]{\oldhref{#1}{\hbox{#2}}}

\title[A precise mass measurement of PSR\,J2045+3633]{A precise mass measurement of PSR\,J2045+3633}

\author[McKee et al.]{J.\,W.\,McKee$^{1,2}$\thanks{E-mail: jmckee@cita.utoronto.ca},
P.\,C.\,C.\,Freire$^{1}$,
M.\,Berezina$^{1}$,
D.\,J.\,Champion$^{1}$,
I.\,Cognard$^{3,4}$,
\newauthor
E.\,Graikou$^{1}$,
L.\,Guillemot$^{3,4}$,
M.\,J.\,Keith$^{5}$,
M.\,Kramer$^{1,5}$,
A.\,G.\,Lyne$^{5}$,
B.\,W.\,Stappers$^{5}$,
\newauthor
T.\,M.\,Tauris$^{6}$,
and G.\,Theureau$^{3,4,7}$
\\
$^{1}$Max-Planck-Institut f{\"u}r Radioastronomie, Auf dem H{\"u}gel 69, D-53121 Bonn, Germany\\
$^{2}$Canadian Institute for Theoretical Astrophysics, University of Toronto, 60 St. George Street, Toronto, ON M5S 3H8, Canada\\
$^{3}$Laboratoire de Physique et Chimie de l'Environnement et de l'Espace LPC2E CNRS-Universit\'{e} d'Orl\'{e}ans, F-45071 Orl\'{e}ans, France \\
$^{4}$Station de Radioastronomie de Nan{\c c}ay, Observatoire de Paris, CNRS/INSU, F-18330 Nan{\c c}ay, France\\
$^{5}$Jodrell Bank Centre for Astrophysics, School of Physics and Astronomy, The University of Manchester, Manchester M13 9PL, UK\\
$^{6}$Department of Physics and Astronomy, Aarhus University, Ny Munkegade 120, 8000 Aarhus C, Denmark\\
$^{7}$Laboratoire Univers et Th\'{e}ories LUTh, Observatoire de Paris, PSL Research University, CNRS/INSU, Universit{\'e} Paris Diderot, 5 place  \\ \ \ Jules Janssen, 92190 Meudon, France \\
}

\date{Accepted XXX. Received YYY; in original form ZZZ}

\pubyear{2020}

\begin{document}
\label{firstpage}
\pagerange{\pageref{firstpage}--\pageref{lastpage}}
\maketitle

\begin{abstract}
We present the results of a timing analysis undertaken with the goal of obtaining an improved mass measurement of the recycled pulsar J2045+3633. Using regular high-cadence observations with the Effelsberg, Nan{\c c}ay, and Lovell radio telescopes, together with targeted campaigns with the Arecibo Telescope and Effelsberg, we have assembled a 6-yr timing data set for this pulsar. We measure highly significant values for the proper motion and the related rate of change of orbital semi-major axis ($\dot{x}$), and have obtained high precision values of the rate of advance of periastron time ($\dot{\omega}$), and two of the Shapiro delay parameters ($h_{3}$ and $\varsigma$). This has allowed us to improve the measurements of the pulsar and companion masses by an order of magnitude, yielding (with $1\sigma$ uncertainties)
$1.251^{+0.021}_{-0.021}\,\text{M}_{\odot}$ for PSR\,J2045+3633, and $0.873^{+0.016}_{-0.014}\,\text{M}_{\odot}$ for its white dwarf companion, and has allowed us to place improved constraints on the geometrical orientation of the binary system. Using our measurements of the binary component masses and the orbital size, we consider possible evolutionary scenarios for the system.
\end{abstract}



\section{Introduction}

\subsection{Binary pulsar timing} \label{subsec_binary_timing}
The timing of radio pulsars, fast-spinning neutron stars (NS) which emit a periodic train of radio pulses, is a powerful tool with a great variety of applications (see examples in \citealp{lk05}). For pulsars in binary systems, precise and continued
timing has allowed for stringent tests of gravity theories (see e.g. \citealt{wex14} for a review),
and detailed studies of the properties of NSs, in particular their masses which are important for the study of
super-dense matter in their interiors (e.g. \citealt{of16}). 
This is possible as pulsar timing analyses of orbits can yield extremely precise measurements of the five Keplerian orbital parameters, measured from the radial motion of the pulsar and in the case of compact systems with a degenerate companion, small relativistic effects on the orbits and the propagation of the radio waves to the observer (\citealp{lk05}, and see \citealp{lor08} for a review). 
These relativistic perturbations (together with other non-Keplerian effects that arise from classical mechanics) can be parameterised in a theory-independent way by a set of
``Post-Keplerian'' (PK) parameters \citep{dt92}. 
The detection of two PK parameters generally allows, under the assumption of a relativistic theory of gravity,
the measurement of the masses of the components of the binary system, and the measurement of additional PK parameters
allows a self-consistency test of that gravity theory (see \citealp{sta03} for a review).

\begin{table*}
\caption{Summary of the configurations used in our observations of PSR\,J2045+3633.}
\label{tab:observationsummary}
\centering 
\begin{tabular} {c c c c c c }
\hline
Telescope & Backend & Centre Frequency (MHz) & Bandwidth (MHz) & Date Range & MJD Range \\
\hline
\hline
Effelsberg (single-pixel receiver) & PSRIX & 1347.5 & 200 & 2014-12-05 $-$ 2016-12-13 & 56996 $-$ 57735 \\
Effelsberg (multi-beam receiver) & PSRIX & 1397.5 & 400 & 2017-05-13 $-$ 2019-09-13 & 57886 $-$ 58739 \\
Nan{\c c}ay & NUPPI & 1484 & 512 & 2015-03-16 $-$ 2020-07-13 & 57097 $-$ 59043 \\
Lovell & ROACH & 1520 & 384 & 2014-09-14 $-$ 2020-02-23 & 56914 $-$ 58902 \\
Arecibo (first campaign) & PUPPI & 1431 & 700 & 2015-08-24 $-$ 2015-09-29 & 57258 $-$ 57294\\
Arecibo (second campaign) & PUPPI & 1381 & 800 & 2019-09-19 $-$ 2019-11-04 & 58745 $-$ 58791\\
\hline
\hline
\end{tabular}
\end{table*}

This technique provided the first indirect detection of gravitational waves (GWs) in the late 1970's, from the
observed orbital decay of the first-discovered binary pulsar, PSR\,B1913+16, the famous ``Hulse-Taylor''
binary (\citealp{wt81}, \citealp{wh16}, \citealp{dam15}, and references therein). Pulsar timing analysis demonstrated that the measured orbital decay was shown to be in exact agreement with
the general relativity (GR) prediction for the energy loss via emission of quadrupolar GWs.
The same technique allowed five independent high-precision tests of GR in the ``double pulsar'' system, using the faster-spinning first-formed pulsar
(PSR\,J0737$-$3039A, \citealt{ksm+06}). Other studies have strongly constrained alternative theories of gravity,
in particular by searching for effects arising from the violation of the strong equivalence principle (SEP). These
include dipolar GW emission (see e.g. \citealp{fwe+12}, and also \citealp{ssb+17}, \citealp{afy19} for recent summaries)
and a violation of the universality of
free-fall (\citealp{agh+18,vfc+20}). No SEP violation has been detected, and to date the results of all high-precision tests of gravity have been consistent with the predictions of GR.

\begin{figure}
	\includegraphics[width=\columnwidth]{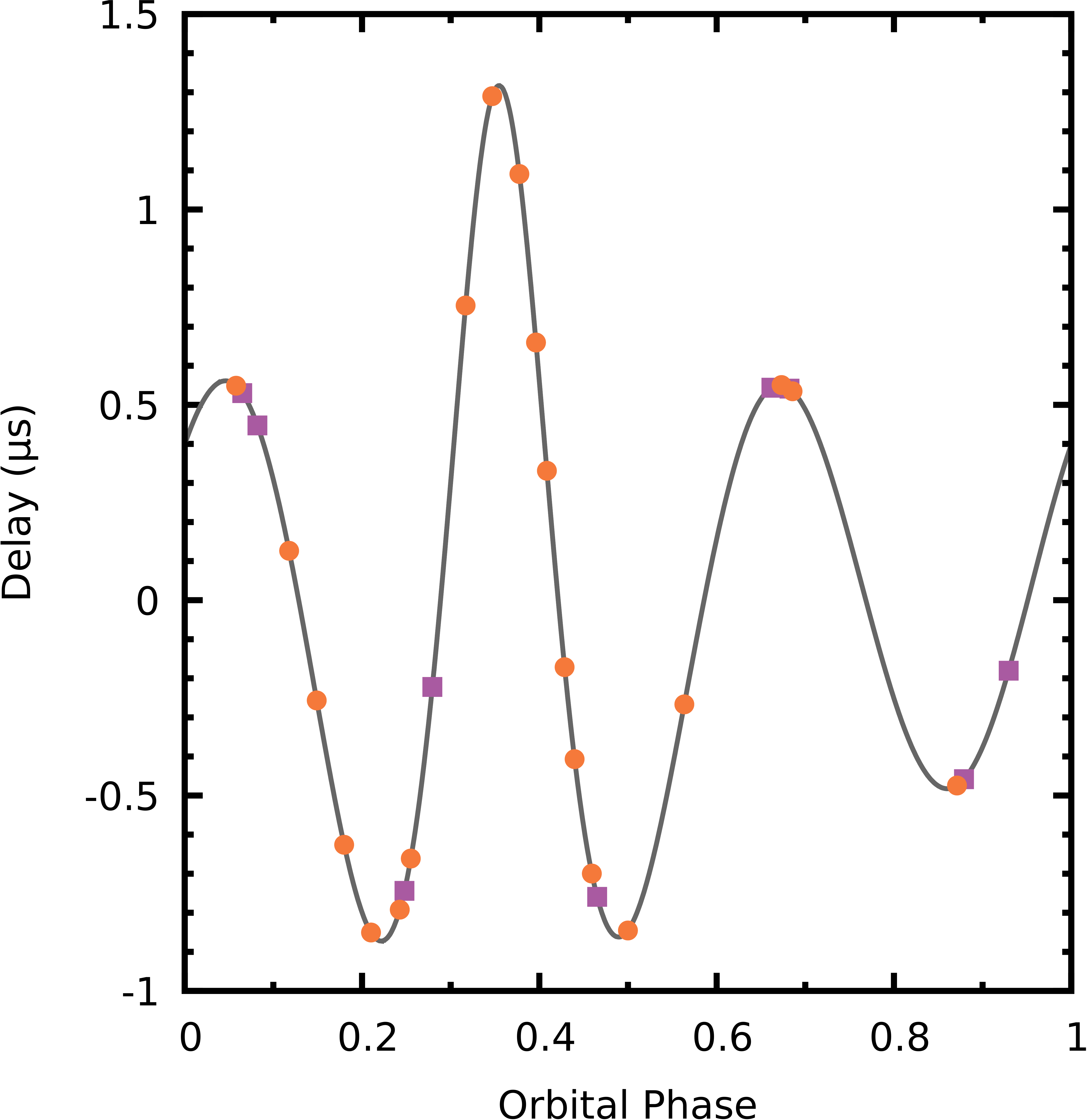}	
	\centering
	\caption{Diagram of the PSR\,J2045+3633 orbital phase relative to periastron showing the harmonic decomposition of the unabsorbed Shapiro delay signal (see text), and highlighting the observations taken during our two targeted campaigns with Effelsberg (orange circles) and Arecibo (purple squares). Observations with both telescopes were scheduled to be close to the local minima and maxima of the Shapiro delay signal, allowing our sensitivity to the overall shape of the signal to be maximised. Longer scans with Effelsberg were used close to the global maximum, to improve our sampling of this part of the orbit.}
	\label{fig:shapiro_delay_sampling}
\end{figure}

This is useful for our purposes, because if we are confident that a particular binary pulsar
has a degenerate companion, then in general classical Newtonian perturbations to the Keplerian orbital
motion are absent. In this case, we can infer the component masses from only two PK parameters, by assuming that
those parameters are as given by GR.

However, measuring two PK parameters is not trivial: at the time of writing,
there are more than 300 known binary pulsars, of which only 38
have precise mass measurements based on this technique\footnote{We define ``precise'' as a pulsar mass $m_{\text{p}}$ measured to 15\% relative uncertainty. See \url{https://www3.mpifr-bonn.mpg.de/staff/pfreire/NS_masses.html}}.
This is because the measurement of PK parameters depends on several factors, some of which are under our control,
such as the length of the timing baselines and, to a lesser extent, the timing precision (which can be improved through
the use of high-gain telescopes, the use of coherent dedispersion, the choice of optimal bands for observing, long integration times, and large bandwidths). Other factors are totally beyond our control, in particular the intrinsic characteristics
of the system: the flux density of the pulsar, its orbital period, eccentricity, inclination, and companion mass.
All of these factors determine the magnitude and measurability of the PK parameters. For eclipsing binary pulsars in
``black widow'' or ``redback'' systems, relativistic perturbations are overwhelmed by the Newtonian effects in the orbits
(e.g. \citealt{svf+16}), preventing the measurement of any PK parameters even under the best conditions.

Although the sample size is small compared to the known population, the number of NS mass measurements already yields impressive results.
Some NS masses are measured to very high precision: e.g. PSR\,B1534+12 has a pulsar mass $m_{\text{p}}=1.3330(2)\,\text{M}_{\odot}$ and a companion mass $m_{\text{c}}=1.3455(2)\,\text{M}_{\odot}$, i.e. a precision $\sigma_{m}/m<0.015\%$ \citep{fst14}. Pulsars also show a range of masses much wider
than thought until only a few years ago, from $1.174(4) \, \text{M}_{\odot}$ \citep{msf+15} to $2\,\text{M}_{\odot}$
and above \citep{afw+13,cfr+20}.

\subsection{Motivation and structure of this work}

Increasing the number of precisely-measured NS systems is useful for several reasons, which we consider here. First,
statistical analyses of the distribution of masses throughout the known population \citep{ato+16} suggest
a bi-modal distribution, but more high-precision measurements are necessary
to either confirm or disprove this finding. Second, precise NS masses are important for astrophysical
studies, in particular the physics of supernovae. An important example of this is the apparent relation between the
supernova kick and the mass of the resulting NS \citep{tkf+17}. Third, the
maximum NS mass represents a fundamental constraint on the equation of state of neutron matter at densities
above those of nuclear matter, a fundamental question in nuclear physics (\citealp{lp01}, \citealp{of16}).

\begin{table*}
\caption{Summary of our timing data set used in this work. The final column refers to the weighted RMS of the post-fit timing residuals with the DDGR model.}
\label{tab:datasummary}
\centering 
\begin{tabular} {c c c c c c }
\hline
Data set  & TOA integration time (s) & TOA Bandwidth (MHz) & $N_{\text{TOA}}$ & EFAC & $W_{\text{RMS}}$ (\SI{}{\micro\second}) \\
\hline
\hline
Effelsberg (single-pixel receiver)  & 600 & 100 & 166 & 1.042 & \ \ 5.57 \\
Effelsberg (multi-beam receiver)  & 600 & 100 & 1016 & 1.069 & \ \ 6.79 \\
Nan{\c c}ay  & 600 & 128 & 1672 & 1.358 & 11.13 \\
Lovell  & 600 & 100 & 1966 & 1.018 & 16.63 \\
Arecibo (first campaign)  & 600 & 100 & 196 & 1.920 & \ \ 2.23 \\
Arecibo (second campaign)  & 600 & 100 & 343 & 1.150 & \ \ 1.79 \\
\hline
All &  - & - & 5359 & - & \ \ 5.52 \\
\hline
\hline
\end{tabular}
\end{table*}

The system described in detail in this work, 
PSR\,J2045+3633, is a 31.7\,ms pulsar discovered by \cite{bcf+17} as part of the
High Time Resolution Universe-North pulsar survey (HTRU-North, \citealp{bck+13}).
The measured spin period $P$ and spin-down rate indicate that this pulsar is old (characteristic age $\tau_{\text{c}}=P/2\dot{P}\sim800$\,Myr), and is therefore part of the ``recycled'' population of pulsars, which were spun up by accretion of matter from another star, the progenitor
of their present companions (described in \citealp{alp82}).
In the same way as most (but not all) recycled pulsars, it is in a binary system, with an orbital period of 32.3 days. From initial analysis following the discovery of the pulsar, the large mass function ($f \, = \, 0.10646 \, \text{M}_{\odot}$) implied that the companion is relatively massive. One of the unusual features of this system is its orbital eccentricity, $e \, \sim \, 0.017$. This is too small for the companion to be a second NS, as the sudden mass loss via the supernova in which it formed would have resulted in a highly-eccentric orbit (see e.g. \citealt{tkf+17}), meaning the companion is more likely a white dwarf (WD). The companion mass is among the largest-known for pulsar-WD systems.
At the time of writing, there has been no attempt to detect the WD companion at optical wavelengths, but the large distance ($\sim5.5$\,kpc) suggests that this would be difficult, while the combination of the large mass and old age imply that the WD has likely cooled to the extent that the brightness is below detectability limits.

Since its discovery, the characteristics of this system have made it a promising candidate for precise mass determination, and the combination of the large companion mass and high timing precision enabled \cite{bcf+17} to detect the Shapiro delay \citep{sha64}.
The large orbital eccentricity also allowed the rate of advance of periastron ($\dot{\omega}$) to be measured and, assuming the validity of GR, they were able to combine both effects to measure masses of 
$1.33^{+0.30}_{-0.28}\,\text{M}_{\odot}$ for the pulsar and $0.94^{+0.14}_{-0.13}\,\text{M}_{\odot}$ for the companion.
These values were not precise enough to be interesting from an astrophysical point of view, however they
made it clear that continued timing would yield much improved measurements of the PK parameters and
therefore much more precise masses for the pulsar
and the companion. In particular, with the Arecibo observations, they were able to achieve a timing precision well under \SI{1}{\micro\second} for the measurement of the topocentric pulse times of arrival (TOAs).

In the remainder of the paper, we will first (Section~\ref{sec:obs}) describe the observations of
PSR\,J2045+3633 used in this work, and how they were analysed. In Section~\ref{sec:results}, we will present our main timing results.
We then discuss their astrophysical implications in Section~\ref{sec:discussion}.
Finally, we summarise our findings in Section~\ref{sec:conclusions}.

\section{Observations and Data Reduction}
\label{sec:obs}

Our data set comprises the observational data that were used in \cite{bcf+17}, as well as data obtained from continued monitoring and special campaigns undertaken for this study. The observations used in \cite{bcf+17} were made with
the Effelsberg Radio Telescope in Germany, The Nan{\c c}ay Radio Telescope in France, and the Lovell Telescope at Jodrell Bank in the United Kingdom, in addition to a dense orbital campaign undertaken with the Arecibo Telescope in the United States.

For this work, we have included an additional four years of data from regular monitoring with the Effelsberg, Nan{\c c}ay, and Lovell telescopes, at cadences of approximately 1 month, 14 days, and 10 days respectively. We also include data taken during two special campaigns: one with Effelsberg and one with Arecibo, conducted with a view to obtaining a precise measurement of the Shapiro delay. The observational setup of the telescopes used in this work is summarised in Table \ref{tab:observationsummary}.

At the Effelsberg telescope, PSR\,J2045+3633 is observed approximately monthly as part of an ongoing key science project to monitor a number of relativistic pulsar binaries. Observations with the Effelsberg telescope used two receivers: a single-pixel receiver with a centre frequency of 1347.5\,MHz and 200\,MHz of bandwidth (observations prior to 2017), and a multi-beam receiver with a centre frequency of 1397.5\,MHz and 400\,MHz of bandwidth (2017 onward). In the case of both receivers, data recording used the ROACH-based PSRIX backend with coherent-dedispersion, detailed in \cite{lkg+16}. In addition to the regular monitoring, a special campaign took place between July and August 2019, and used scans lasting between 1.5$-$3.5\,hrs, totalling $\sim45$\,hrs. A higher cadence of observations was used at orbital phases corresponding to local maxima and minima of the harmonic decomposition of the unabsorbed Shapiro delay signal, calculated from Equation 28 of \cite{fw10}, with particular focus on the global maximum, with the goal of placing better constraints on the Shapiro delay signal, rather than when solely focusing on the maximum delay corresponding to the superior conjunction (Figure \ref{fig:shapiro_delay_sampling}).

\begin{figure}
	\includegraphics[width=\columnwidth]{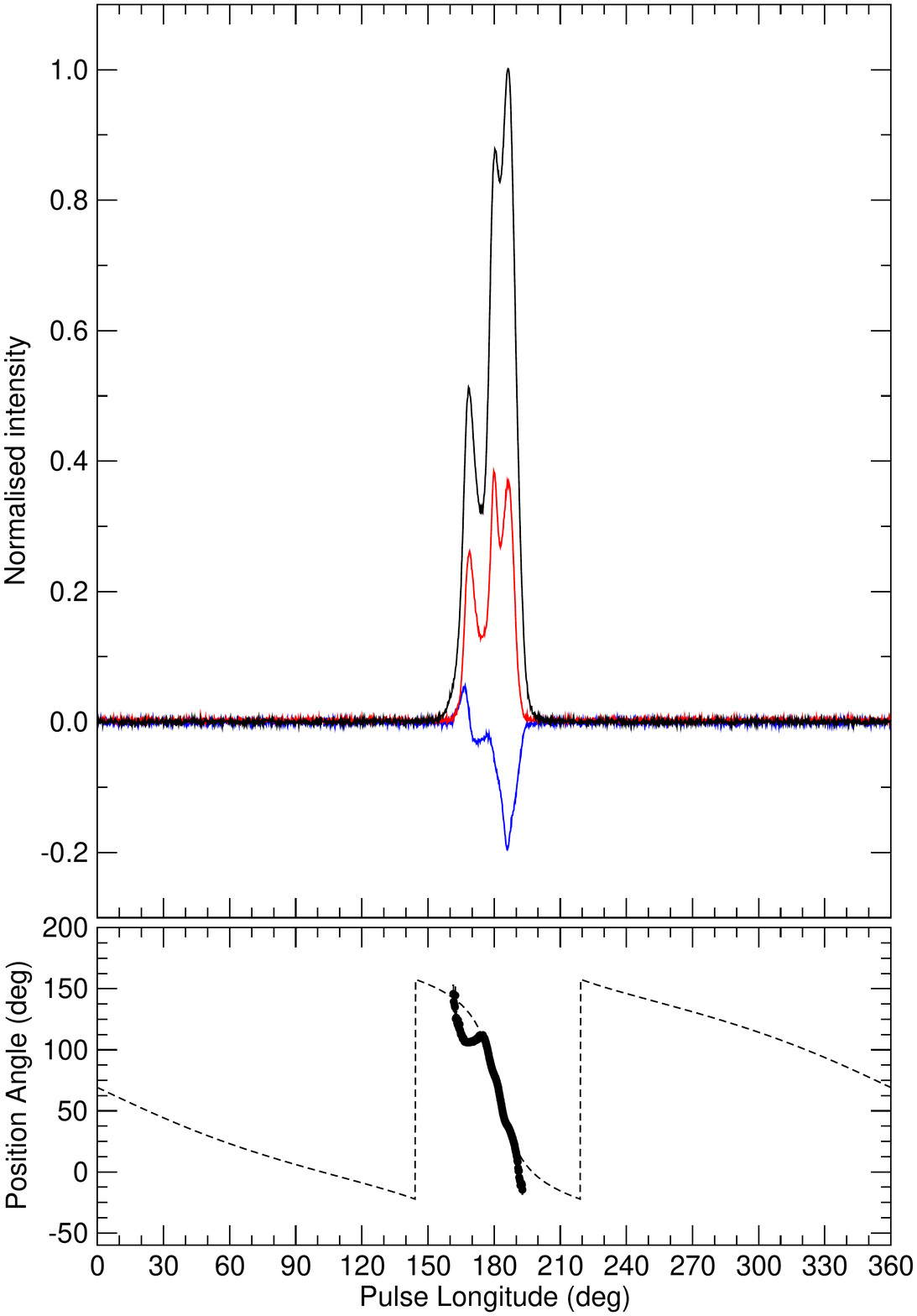}	
	\centering
	\caption{The pulse profile of PSR\,J2045+3633, obtained from a 62-minute scan with the Arecibo Telescope at 1431\,MHz (700\,MHz bandwidth). The observational data were polarisation-calibrated, corrected for Faraday rotation, and then fully integrated in time and observing frequency. The top panel shows the total intensity (Stokes $I$, black line), the corresponding linearly-polarised intensity (Stokes $L$, red line), and the circularly-polarised intensity (Stokes $V$, blue line). The S/N of this observation is approximately 3000. The lower panel shows the position angle (PA) of the linearly polarised emission. The dashed line shows a modeling of the PA values within the Rotating Vector Model (RVM, see text for details.)}
	\label{fig:J2045polprof}
\end{figure}

The Lovell observations were taken using a receiver working in the frequency range 1300--1700\,MHz, with a maximum usable bandwidth of 400\,MHz, and acquired using a ROACH system
detailed in \cite{bjk+16}. Observations of PSR\,J2045+3633 are taken approximately every 2 weeks, as part of the Jodrell Bank monitoring programme of $\sim800$ pulsars in the Northern Sky. 

The Nan{\c c}ay Radio Telescope observations used a 1484\,MHz receiver with 512\,MHz of bandwidth, and recorded with a dedispersing ROACH backend (NUPPI\footnote{\url{https://github.com/gdesvignes/NUPPI}}), which is described in \cite{ldc+14}. PSR\,J2045+3633 is observed approximately every 10 days with the Nan{\c c}ay Radio Telescope, as part of a campaign to provide high-cadence monitoring of $>100$ recycled pulsars.

\begin{figure*}
    \advance\leftskip-0.12cm
	\includegraphics[scale=0.38]{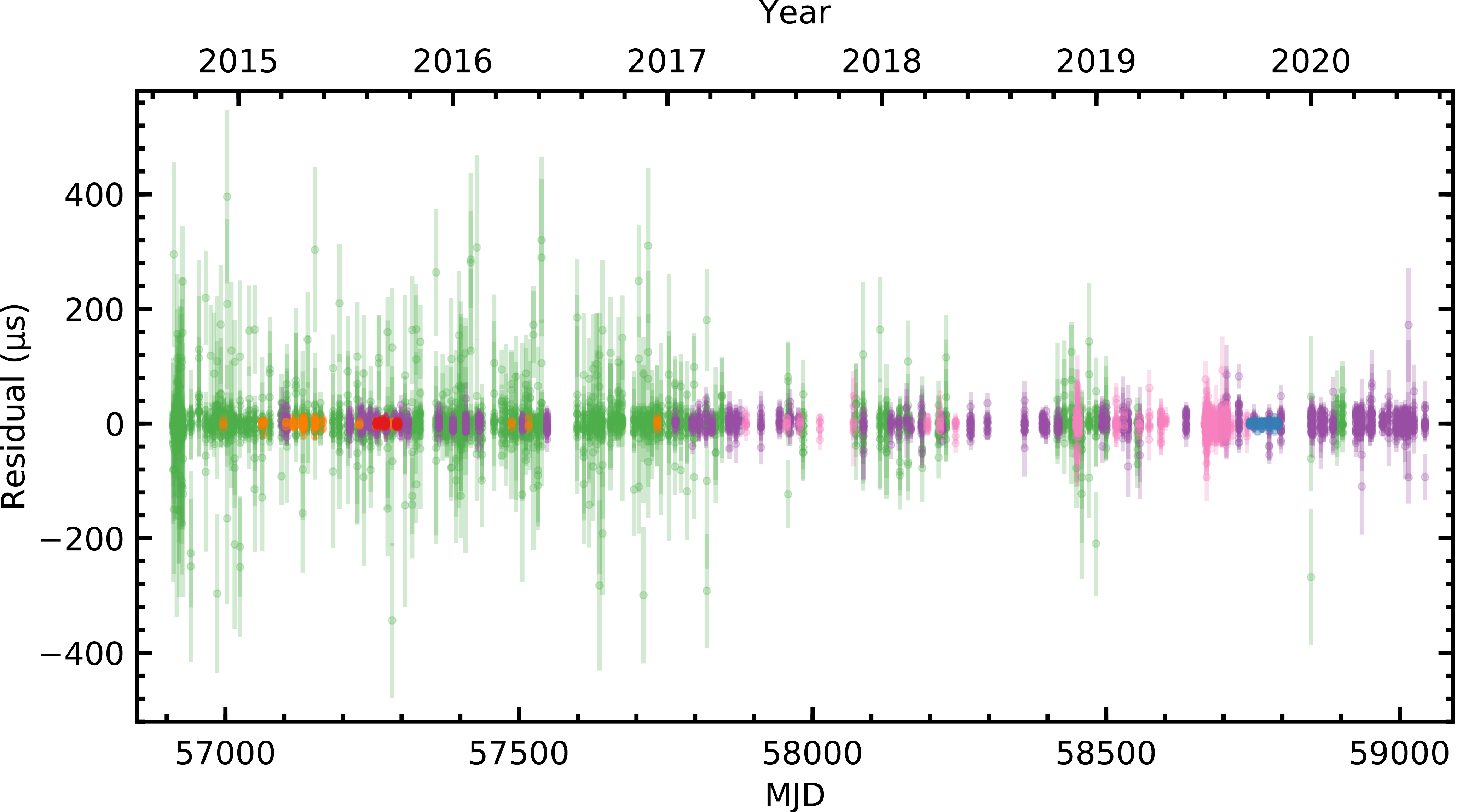}	
	
	\vspace{0.5cm}
	
	\includegraphics[scale=0.38]{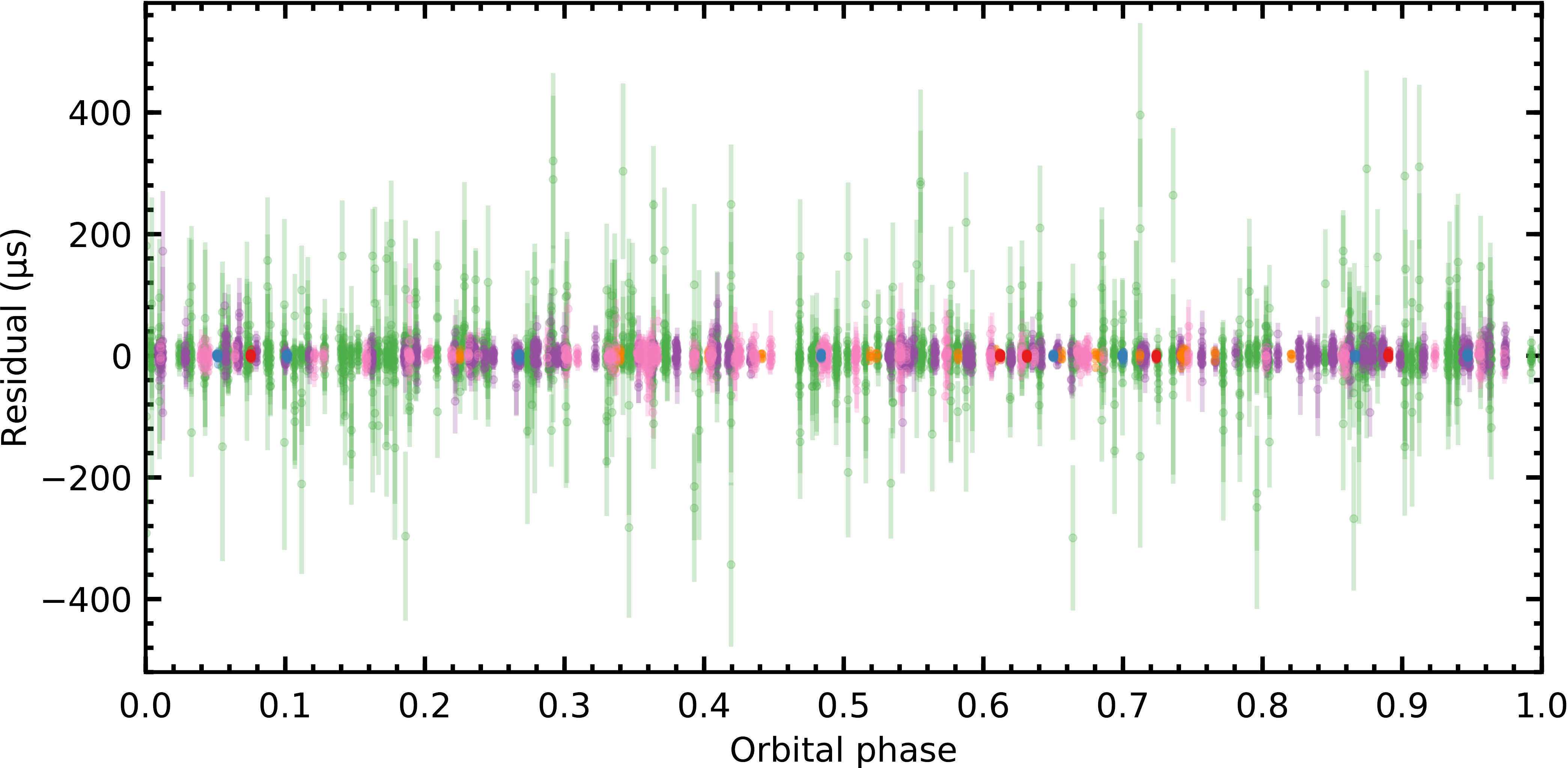}	
	\centering
	\caption{PSR\,J2045+3633 timing residuals, plot as a function of MJD (above) and orbital phase relative to periastron (below), and analysed using the DDGR orbital model (see text). Data up to MJD\,57538 were included in \protect\cite{bcf+17}. The weighted RMS of the timing residuals is \SI{5.52}{\micro\second}. The following colour coding is used: Lovell (green), Nan{\c c}ay Radio Telescope (purple), Effelsberg Telescope single-pixel (orange), Effelsberg Telescope multi-beam (pink), Arecibo Telescope 1431\,MHz (red), Arecibo Telescope 1381\,MHz (blue).}
	\label{fig:J2045residuals}
\end{figure*}

Our Arecibo Telescope data set is made up of two targeted campaigns. The first was taken in late 2015, and is presented in \cite{bcf+17}. Observations from the first campaign used a centre frequency of 1431\,MHz and a bandwidth of 700\,MHz, and were coherently-dedispersed. The second campaign was undertaken specifically for this work, and took place between September and November 2019. 
The observation dates were chosen to sample the turning points of the harmonic decomposition of the Shapiro delay signal (although not the maximum, due to missed observations), which we illustrate in Figure \ref{fig:shapiro_delay_sampling}.
A centre frequency of 1381\,MHz and a bandwidth of 800\,MHz was used. Observations used scans of approximately 1\,hr in length, with the entire campaign totalling $\sim13$\,h. Figure~\ref{fig:J2045polprof} shows the pulse profile obtained during one such observing scan.

The observations from all telescopes were refolded with the same ephemeris (adapted from the one presented in \citealp{bcf+17}), and coherently dedispersed with the same dispersion measure (DM). In the case of Effelsberg, Nan{\c c}ay, and Arecibo, polarisation calibration was applied, using noise diode scans that took place immediately before the observations, of lengths 2--5\,min (Effelsberg and Arecibo) and 10\,s (Nan{\c c}ay). The data were also corrected for rotation measure, using the value measured by \cite{bcf+17}. Observations were manually inspected for RFI, with channels and sub-integrations badly-affected by RFI being masked, using the \textsc{pazi} tool from \textsc{psrchive}\footnote{\url{http://psrchive.sourceforge.net/}} \citep{hvm04}.

Often in high-precision timing studies of pulsars, TOAs are generated for several sub-bands across the total observing bandwidth, with the purpose of measuring and removing the dispersive delay due to the ISM (e.g. \citealp{aab+20}), or to give greater weighting to parts of the band that have been enhanced in flux by interstellar scintillation. Until recently, this approach would typically use a single template, created from the frequency-averaged total bandwidth, to generate a TOA from each frequency channel. Some pulsars, including PSR\,J2045+3633, display significant profile evolution across wide observing bandwidths, meaning that a single template will not optimally-describe the pulse shape in each channel. In this case, an additional set of parameters (known as FD parameters) are required to be fit in the timing model to account for the additional fixed phase delays per channel (see e.g. \citealp{aab+20}). In our analysis, we instead use frequency-resolved templates to improve timing precision when compared to the single-template case, and to remove the requirement of fitting a set of FD parameters in the timing model.

\begin{table*}
\caption{PSR\,J2045+3633 spin and astrometric parameters measured from our timing analysis, and associated quantities derived from them. \newline $^{\textit{a}}$ Obtained from \protect\cite{bcf+17} \newline
$^{\textit{b}}$ Assuming a $20\%$ uncertainty \newline
$^{\textit{c}}$ Assuming DM-derived distance (NE2001)}
\label{tab:timing_values}
\centering 
\begin{tabular} {l c}
\hline
\hline
\multicolumn{2}{c}{Observation and data reduction parameters}\\
\hline
Solar system ephemeris\dotfill & DE435 \\
Reference epoch (MJD)\dotfill & 56855.0 \\
Solar wind electron number density, $n_{0}$ (cm$^{-3}$)\dotfill & 7.0 \\
\hline
\multicolumn{2}{c}{Spin and astrometric parameters}\\
\hline
Right ascension, $\alpha$ (J2000, h:m:s)\dotfill & 20:45:01.50510(2)  \\
Declination, $\delta$ (J2000, d:m:s)\dotfill & 36:33:01.4046(2)  \\
Proper motion in $\alpha$, $\mu_{\alpha}$ (mas\,yr$^{-1}$)\dotfill & $-3.20(6)$  \\
Proper motion in $\delta$, $\mu_{\delta}$ (mas\,yr$^{-1}$)\dotfill  & $-2.68(6)$ \\
Spin frequency, $\nu$ (Hz)\dotfill & 31.563820566264(3)  \\
Spin-down rate, $\dot{\nu}$ ($10^{-16}$\,Hz\,s$^{-1}$)\dotfill & $-5.8628(3)$  \\
Dispersion measure, DM (cm$^{-3}$\,pc)\dotfill  & 129.533(2) \\
First derivative of DM, DM1 ($10^{-3}$\,cm$^{-3}$\,pc\,yr$^{-1}$)\dotfill  & $+6.1(16)$ \\
. . . DM2 ($10^{-3}$\,cm$^{-3}$\,pc\,yr$^{-2}$)\dotfill  & $-7.6(7)$ \\
. . . DM3 ($10^{-3}$\,cm$^{-3}$\,pc\,yr$^{-3}$)\dotfill  & $+3.8(2)$ \\
Rotation measure, RM (rad\,m$^{-2}$)\dotfill  & $-266(10)^{\textit{a}}$ \\
\hline
\multicolumn{2}{c}{Derived parameters}\\
\hline
Galactic longitude, $l$ ($^{\circ}$)\dotfill  & 77.8323570(2) \\
Galactic latitude, $b$ ($^{\circ}$)\dotfill  & $-3.92576397(7)$ \\
Total proper motion, $\mu_{\text{T}}$ (mas\,yr$^{-1}$)\dotfill  & 4.17(6) \\
Position angle of proper motion (J2000), $\Theta_{\mu}$ ($^{\circ}$) \dotfill & 230.1(8) \\
Position angle of proper motion (Galactic), $\Theta_{\mu}$ ($^{\circ}$) \dotfill & 281.2(8)  \\
DM-derived distance (NE2001), $d$ (kpc)\dotfill & $5.5(11)^{\textit{b}}$  \\
DM-derived distance (YMW16), $d$ (kpc)\dotfill & $5.6(11)^{\textit{b}}$  \\
Parallax, $\bar{\omega}$ (mas)\dotfill  & 0.18$^{\textit{c}}$ \\
Galactic height, $z$ (kpc)\dotfill & $-0.381^{\textit{c}}$  \\
Heliocentric transverse velocity, $v_{\text{T}}$ (km\,s$^{-1}$)\dotfill & $109(22)^{\textit{c}}$  \\
Spin period, $P_{0}$ (ms)\dotfill & 31.681842757300(3)\\
Spin period derivative, $\dot{P}$ ($10^{-19}$\,s\,s$^{-1}$)\dotfill & 5.8847(3) \\
Total kinematic contribution to period derivative, $\dot{P}_{\text{k}}$ ($10^{-19}$\,s\,s$^{-1}$)\dotfill & $-0.074$  \\
Intrinsic spin period derivative, $\dot{P}$ ($10^{-19}$\,s\,s$^{-1}$)\dotfill & $5.953^{+0.001}_{-0.005}$ \\
Surface magnetic field strength, $B_{\text{surf}}$ ($10^{9}$\,G)\dotfill  & 4.4 \\
Characteristic age, $\tau_{\text{c}}$ (Gyr)\dotfill & 0.84  \\
Spin-down power, $\dot{E}$ ($10^{32}$\,erg\,s$^{-1}$)\dotfill & 7.4  \\
\hline
\hline
\end{tabular}
\end{table*}

\begin{table*}
\caption{PSR\,J2045+3633 binary parameters measured from our timing analysis, with values separated by orbital model (see text). Square brackets indicate derived quantities that are not directly measured. The results of the $\chi^{2}$ mapping produce degenerate solutions for the orbital geometry, and so two values for $i$ and four for $\Omega$ are listed.}
\label{tab:binary_results}
\centering 
\begin{tabular} {l c c c }
\hline
\hline
Binary model\dotfill & DDGR & DDFWHE  & DDK $\chi^{2}$ grid \\[0.5ex]
\hline
\multicolumn{4}{c}{Keplerian orbital parameters}\\[0.5ex]
\hline
Orbital period, $P_{\text{b}}$ (days)\dotfill & 32.29784447(8) & 32.29784448(8) & - \\[0.5ex]
Projected semi-major axis of the pulsar orbit, $x$ (lt-s)\dotfill & 46.940800(5) & 46.940797(4) & - \\[0.5ex]
Epoch of periastron, $T_{0}$ (MJD)\dotfill & 57496.750854(3) & 57496.750850(7) & - \\[0.5ex]
Orbital eccentricity, $e$\dotfill & 0.017212447(6) & 0.01721245(2)  & - \\[0.5ex]
Longitude of periastron, $\omega$ ($^{\circ}$)\dotfill & 320.77788(3) & 320.77783(8) & - \\[0.5ex]
Orbital inclination, $i$ ($^{\circ}$)\dotfill & [63] & 64(7) & $63.8^{+1.5}_{-1.6}$, $117.5^{+1.6}_{-1.5}$ \\[0.5ex]
Position angle of the orbital line of nodes, $\Omega$ ($^{\circ}$)\dotfill & - & - & $9^{+13}_{-20}$, $92^{+16}_{-13}$,  $187^{+13}_{-18}$, $272^{+16}_{-13}$ \\[0.5ex]
\hline
\multicolumn{4}{c}{Post-Keplerian orbital parameters}\\[0.5ex]
\hline
Shapiro delay `shape', $s$\dotfill & [0.8932797] & 0.90(6) & - \\[0.5ex]
Rate of advance of periastron, $\dot{\omega}$ ($^{\circ}$\,yr$^{-1}$)\dotfill & [0.0010074] & 0.001009(10) & - \\[0.5ex]
Orbital period derivative, $\dot{P_{\text{b}}}$ ($10^{-12}$\,s\,s$^{-1}$)\dotfill & 3.5(53) & 3.6(53) & - \\[0.5ex]
Rate of change of orbital semi-major axis, $\dot{x}$ ($10^{-15}$\,lt-s\,s$^{-1}$)\dotfill & $-9.6(28)$ & $-8.7(28)$ & - \\[0.5ex]
Orthometric amplitude of the Shapiro delay, $h_{3}$ (\SI{}{\micro\second})\dotfill & - & 1.01(11) & - \\[0.5ex]
Orthometric ratio of the Shapiro delay, $\varsigma$\dotfill & - & 0.62(9) & - \\[0.5ex]
\hline
\multicolumn{4}{c}{Mass measurements}\\
\hline
Mass function, $f$ (M$_{\odot}$)\dotfill & 0.10646016(3) & 0.10646014(3) & - \\[0.5ex]
Total mass, $M$ (M$_{\odot}$)\dotfill & 2.14(3) & [2.1353] & $2.127^{+0.031}_{-0.031}$ \\[0.5ex]
Pulsar mass, $m_{\text{p}}$ (M$_{\odot}$)\dotfill & 1.26(3) & [1.2453] & $1.251^{+0.021}_{-0.021}$ \\[0.5ex]
Companion mass, $m_{\text{c}}$ (M$_{\odot}$)\dotfill & 0.88(1) & 0.9(4) & $0.873^{+0.016}_{-0.014}$ \\[0.5ex]
\hline
\hline
\end{tabular}
\end{table*}

For our frequency-resolved timing, we followed the approach detailed by \cite{dvt+19}, who employed this technique on LOFAR and GLOW observations of pulsars at frequencies $<150$\,MHz, where intrinsic profile evolution with frequency is very rapid, and extrinsic frequency-dependent effects such as scattering and DM variations are highly significant. We made high-S/N reference profiles for each telescope by summing the top $10\%$ highest-S/N observations from each, after fully integrating in observation time and polarisation, to increase S/N and reduce correlated noise. These reference templates were then integrated to have 100\,MHz channel widths in the case of Effelsberg, Lovell, and Arecibo, and 128\,MHz in the case of Nan{\c c}ay. A wavelet smoothing algorithm was then applied to each channel, to further increase S/N. We then integrated our data sets in frequency so that the same channel widths as the templates were used, and with 10-min sub-integration times. From these, TOAs were generated by cross-correlating the observational data with the frequency-resolved reference templates, using a Fourier domain MCMC algorithm included in the \textsc{pat} tool in \textsc{psrchive}. Our timing data set is summarised in Table \ref{tab:datasummary}.

Our timing analysis follows the standard techniques that have been used extensively in studies such as this (see e.g. \citealp{dcl+16}). The TOAs from each unique telescope and receiver system were aligned by fitting for time offsets. DM variations were modelled using a Taylor series, with higher-order terms being added until the $\chi^{2}$ of the resulting fit was reduced by less than one, which resulted in a 3$^{\text{rd}}$ order Taylor series for the final DM model. For each of these data sets, the uncertainties were scaled by a constant factor (``EFAC'' in Table \ref{tab:datasummary}) such that the post-fit reduced $\chi^{2}=1$. We used the \textsc{tempo}\footnote{\url{http://tempo.sourceforge.net/}} pulsar timing software for our analysis, and we present our timing residuals (i.e. the difference between observed TOAs and our timing model) in Figure \ref{fig:J2045residuals}. 

Our analysis of the binary parameters started from the timing model presented in \cite{bcf+17}.
This was adapted to derive new timing models, which are extensions of the theory-independent DD model (\citealp{dd85}, \citealp{dd86}). The first is the DDGR model, which assumes the validity of GR and fits self-consistently for the
masses of the components in the binary system. The second is the DDFWHE model, which includes the orthometric re-parameterisation of the Shapiro delay
\citep{fw10}; this was implemented in {\sc tempo} by \cite{wh16}. Finally, the 
DDK model (extended to include kinematic effects, \citealp{kop95}, \citealp{kop96}) was used to
calculate $\chi^2$ maps of the binary parameter space (discussed in the next section). All analyses used the DE435 planetary ephemeris \citep{fjp+16} to correct for the motion of the radio telescope relative to the Solar System barycentre.

\begin{table*}
\caption{Results of the $\chi^{2}$ mapping with the DDK model for the two $\cos i$ regions (see text).}
\label{tab:gridresults}
\centering 
\begin{tabular} {c c c c c }
\hline
$\cos i$ range & Best $\cos i$ & Best $\Omega$ ($^{\circ}$) & Best $M$ ($\text{M}_{\odot}$) & Minimum $\chi^{2}$ \\
\hline
$-0.9:-0.1$ & $-0.45$ & 190 & 2.13 & 5582.83 \\
$+0.1:+0.9$ & \ \ +0.45 & 269 & 2.13 & 5583.75 \\
\hline
\end{tabular}
\end{table*}

\section{Results}
\label{sec:results}

Our timing analysis resulted in the measurements of the timing parameters presented in Tables~\ref{tab:timing_values} and \ref{tab:binary_results}. In the latter Table,
we present the results of two models: one that assumes the validity of GR (the DDGR model), and  a theory-independent model (the DDFWHE model). We also present there the results of the $\chi^2$ mapping
we have carried out using the DDK model. All parameters are expressed in Dynamic Barycentric Time (TDB), the spin and astrometric
parameters refer to the reference epoch
listed, and the orbital parameters are valid for the time of passage through periastron listed ($T_0$). All parameters
are quoted with 1$\sigma$ confidence limits. The timing
residuals (i.e. the measured TOAs minus the DDGR model prediction for their respective rotation numbers) are presented
in Figure \ref{fig:J2045residuals}, which demonstrates that the TOAs are well-described by the models in Tables \ref{tab:timing_values} and \ref{tab:binary_results}. 

In what follows, we present a discussion on the different timing parameters in these tables.

\subsection{Proper motion and distance}

One of the new results from this work is a high-precision measurement of the proper motion PSR\,J2045+3633. We measure the total proper motion to be
4.17(6)\,mas\,yr$^{-1}$, with a position angle (PA) of the proper motion ($\Theta_\mu)$ of
$230.1(8)^{\circ}$ in J2000 coordinates and $281.2(8)^{\circ}$ in Galactic 
coordinates. The convention used here for the PA is the so-called ``observer's convention", where a
PA of $0^{\circ}$ indicates North, and $90^{\circ}$ indicates East.
In Galactic coordinates, the horizontal and vertical components of its motion are given by
$-4.1$ and +0.8\,mas\,yr$^{-1}$, respectively. The first minus sign indicates a Westwards motion in the Galaxy,
i.e. in a sense of decreasing Galactic longitude $l$. The second plus sign means the pulsar is increasing its
Galactic latitude $b$, i.e. it is slowly approaching the Galactic plane.

In order to calculate the heliocentric velocity, we must have an estimate of the distance to the system, $d$.
We do not detect a parallax signal from the pulsar timing,
and we instead use an estimated distance from two Galactic electron density models: NE2001 \citep{cl02} and YMW16 \citep{ymw17}.
These models use the observed DM of a series of pulsars with independent distance measurements
(e.g. parallax from timing or VLBI) to estimate the free electron density along particular lines of site, i.e.
\begin{equation}\label{dispersion_measure}
    \text{DM}=\int_{0}^{d} n_{\text{e}}\,\rm{d}\textit{l} .
\end{equation}
These models only represent a crude estimate, particularly for distant pulsars, where independent distance measurement are difficult to obtain for that part of the Galaxy.
We therefore assign a conservative estimate of the distance uncertainty of $20\%$, take the distance values as $d_{\text{NE2001}} \, = \, 5.5(11)$\,kpc and $d_{\text{YMW16}} \, = \, 5.6(11)$\,kpc. These estimates are mutually consistent.
For the remainder of the calculations, we use the NE2001 estimate with the 20\% distance uncertainty.

The NE2001 distance and its $20\%$ uncertainty implies a Heliocentric velocity of 109(22)\,km\,s$^{-1}$.
This is only a basic estimate of the dynamics of the system, and we will present a
more detailed analysis of the motion of the system in Section~\ref{sec:discussion}.

\subsection{Keplerian parameters and the mass function}

The five Keplerian parameters measured to high precision in all binary pulsar systems are 
the orbital period ($P_{\text{b}}$),
the semi-major axis of the pulsar's orbit projected along the line of sight ($x$),
the orbital eccentricity ($e$), the longitude of periastron ($\omega$), and the time of passage through 
periastron ($T_0$).
The first two yield the mass function from Kepler's third law:
\begin{equation}\label{mass_fn_eqn}
    f(m_{\text{p}},m_{\text{c}})=\frac{(m_{\text{c}}\sin i)^{3}}{M^{2}}=\frac{4\pi^{2}}{T_{\odot}}\frac{x^{3}}{P_{\text{b}}^{2}} = 0.10646016(3)\,\text{M}_{\odot},
\end{equation}
where $M = m_{\text{p}}+m_{\text{c}}$ is the total mass of the system, $m_{\text{p}}$ and $m_{\text{c}}$ are the masses of the pulsar and companion respectively, $i$ is the orbital inclination angle, and $T_{\odot}=G M_{\odot}c^{-3}=\SI{4.925490947641267}{\micro\second}$ is a constant describing the speed of light travel time across a 1\,M$_{\odot}$ gravitational radius (here, $G$ is the gravitational constant, and $c$ is the speed of light; since the factors
$G$, $M_{\odot}$, and $c$ are defined exactly, $T_{\odot}$ is also an exact constant). As $x$  and $P_{\text{b}}$ are usually measurable to high precision, the mass function is often very well known. However, measurements of other quantities are required to disentangle $m_{\text{p}}$ and $m_{\text{c}}$.

\subsection{Rate of advance of periastron} \label{subsec_periastron_advance}

From our timing analysis, we have greatly improved the measurement of the rate of advance of periastron, yielding the value $\dot{\omega}=0.001009(10)^{\circ}\,\text{yr}^{-1}$. In the absence of other nearby (i.e. gravitationally-interacting) objects, this is given by:
\begin{equation}\label{omega_dot_contributions}
    \dot{\omega}_{\text{obs}}=\dot{\omega}_{\text{rel}}+\dot{\omega}_{\text{k}}+\dot{\omega}_{\text{SO}} .
\end{equation}
The first term is the relativistic periastron advance. Assuming that GR is the correct theory of
gravity, this is given by \citep{tw82}:
\begin{equation}\label{omdot_eqn}
    \dot{\omega}_{\text{rel}} \, = \, 3T_{\odot}^{2/3}  \left(\frac{P_{\text{b}}}{2\pi}\right)^{-5/3} \frac{1}{1-e^{2}} M^{2/3}.
\end{equation}
From our measured value of $\dot{\omega}$, we obtain a total mass of $2.14(3) \, \text{M}_{\odot}$.
This constraint on the total mass is represented by the red lines in Figure~\ref{fig:J2045_om_mass}.

The second term, $\dot{\omega}_{\text{k}}$, is the kinematic contribution. This becomes important for several
binary systems that have orbital periods similar to those of PSR\,J2045+3633
(\citealp{fbw+11}, \citealp{sfa+19}), and is given by \citep{kop96}:
\begin{equation}\label{omega_dot_kinematic}
    \dot{\omega}_{\text{k}}\, = \, \frac{\mu_{\text{T}}}{\sin i}\cos(\Theta_{\mu}-\Omega),
\end{equation}
where $\mu_{\text{T}}=\sqrt{\mu_{\alpha}^{2}+\mu_{\delta}^{2}}$ is the total pulsar proper motion (i.e. the magnitude of the proper motion in right ascension and declination components, $\alpha$ and $\delta$), $\Theta_{\mu}$ is the position angle of the proper motion, and $\Omega$ is the position angle of the orbital line of nodes. 
For PSR\,J2045+3633, we can derive maximum and minimum limits (i.e. when the proper motion is orthogonal to the line of nodes) for this effect from the estimate of $i$ in Table~\ref{tab:binary_results} and the measured proper motion. From this, we obtain
$| \dot{\omega}_{\text{k}} | \, \leq \, 1.4 \times \, 10^{-6} {^{\circ}}\,\text{yr}^{-1}$. This
is smaller than the measurement uncertainty of $\dot{\omega}$, which is approximately $10^{-5}  {^{\circ}}\,\text{yr}^{-1}$. This means that the uncertainty on the total mass of the system is still dominated by the measurement uncertainty
for $\dot{\omega}$.

The final term in Equation \ref{omega_dot_contributions} arises from the spin-orbit coupling of the system, and is negligible in the case
of PSR\,J2045+3633. We discuss this in detail in Section~\ref{sec:SO}.

Since $M$ is known, we can immediately estimate the orbital separation $a$ from Kepler's third law:
\begin{equation}
\frac{a}{c} \, = \, \left[ M T_{\odot} \left(  \frac{P_{\text b}}{2 \pi}  \right)^2  \right]^{1/3} \, = \, 127.6(6) \, \text{lt-s}.
\end{equation}
This value is independent of the orbital inclination $i$, and the
uncertainty is dominated by the uncertainty in $M$. We comment on the implications of this value in Section \ref{subsec_binary_evol}.

\subsection{Shapiro delay}
Our dense observations made around superior conjunction have allowed us to make improved measurements of the Shapiro delay parameters in this system. In the DD orbital model, the Shapiro delay is 
characterised by two PK parameters: the range ($r$) and shape ($s$). In GR, these relate to the masses and orbital inclinations
according to the following expressions:
\begin{equation}\label{shapiro_r_eqn}
    r=\frac{Gm_{\text{c}}}{c^{3}}=T_{\odot}m_{\text{c}},
\end{equation}
\begin{equation}\label{shapiro_s_eqn}
    s=\sin i=T_{\odot}^{-1/3}\left(\frac{P_{\text{b}}}{2\pi}\right)x\frac{(m_{\text{p}}+m_{\text{c}})^{2/3}}{m_{\text{c}}} .
\end{equation}
For most systems, particularly those with low orbital inclinations, the $r$ and $s$ parameters are strongly correlated.
For such systems, it is better to use an alternative set of
PK parameters: the orthometric amplitude ($h_{3}$) and ratio ($\varsigma$) respectively \citep{fw10}, to describe the Shapiro delay. These are given by:
\begin{equation}\label{stig_eqn}
    \varsigma=\frac{\sin i}{1+\sqrt{1-\sin^{2}i}}=\tan\left(\frac{i}{2}\right) ,
\end{equation}
\begin{equation}\label{h3_eqn}
    h_{3}=r\varsigma^{3}.
\end{equation}
The orthometric parameters do not yield better mass measurements, but they are less correlated with each other and therefore provide a better
description of the regions of the $\cos i$--$m_c$ plane where the system is most likely to be located.
Our measurements yield highly significant estimates for these parameters: $h_{3}=\SI{1.01(11)}{\micro\second}$ and $\varsigma=0.62(9)$. The GR mass constraints corresponding to these parameters 
are represented by the blue solid and dashed lines  in Figure~\ref{fig:J2045_om_mass}. As we can see there, all
three PK parameters converge at a consistent set of masses, within our measurement uncertainties. This represents, formally, a successful test of
GR, however the low precision of the masses provided by our Shapiro delay alone reduces the interest
of this test.

Our mass measurements imply, using Equation~\ref{mass_fn_eqn}, an orbital inclination of either
$63(2)^\circ$ or $180^\circ - 63(2)^\circ=117(2)^\circ$. Within the uncertainties (discussed in detail below), the first value
is consistent with the estimate of $i \, \sim \, 60^\circ$ made by \cite{bcf+17} for this system based on a rotating vector model (RVM, \citealp{rc69})
fit to polarisation data of the pulsar.  
We have performed a similar RVM fit to the PAs measured as shown in Figure~\ref{fig:J2045polprof}. Due to the much higher S/N of the Arecibo data, the uncertainties in our PA values are much smaller than those of \cite{bcf+17}, and in order to obtain a suitable fit (shown by the dashed line in Figure~\ref{fig:J2045polprof}), we increased the uncertainties by a factor of 10. The result suggests a viewing angle $\zeta=131^{+30}_{-23} {^\circ}$, whereas the large uncertainties are not surprising given the relatively short longitude range that is available for the RVM fit (cf.~\citealt{lk05}). This implies an orbital inclination angle $i=49^{+23}_{-30} {^\circ}$ (see discussion in \citeauthor{ksv+20} submitted), which is both consistent with the result of \cite{bcf+17} and the solution for the Shapiro delay that favours $i=63(2)^\circ$.
\label{sec:rvm}

\subsection{Variation of the projected semi-major axis} \label{subsec_semi_major_ax}

In addition to the large improvements in precision to the parameters reported by \cite{bcf+17}, we have measured a significant variation in the orbital semi-major axis, ${\dot{x}=-8.7(28)\times10^{-15}\,\text{lt-s\,s}^{-1}}$. This allows us to impose a further constraint on the orbital geometry of the system.

The observed $\dot{x}$ is composed of a combination of intrinsic, geometric, and kinematic effects:
\begin{equation}\label{xdot_contributions}
    \left(\frac{\dot{x}}{x}\right)^{\text{obs}}=\left(\frac{\dot{x}}{x}\right)^{\text{k}}+\left(\frac{\dot{x}}{x}\right)^{\text{GW}}+\frac{\rm{d}\epsilon_{\text{A}}}{\rm{d}t}-\frac{\dot{D}}{D}+\left(\frac{\dot{x}}{x}\right)^{\dot{m}}+\left(\frac{\dot{x}}{x}\right)^{\text{SO}} ,
\end{equation}
where the terms in the above equation describe the contributions from the kinematic effects, shrinking of the orbit due to GW emission, aberration, the variation of the Doppler shift, radiative mass loss, and spin-orbital coupling, respectively (these are described in detail in \citealp{lk05}). The spin-orbit coupling term was recently detected for the first time in a double-degenerate system, PSR\,J1141$-$6545 \citep{vbs+20}. We discuss the Newtonian and relativistic
contributions of the spin-orbital coupling in detail in Section~\ref{sec:SO}, and conclude that is is not of importance
for this study.

A detailed analysis of the components of Equation \ref{xdot_contributions} shows that, for PSR\,J2045+3633, only the first term is important.
This term describes the kinematic contribution, arising from the secular change of the orbital inclination angle due to the proper motion of the pulsar \citep{kop96}:
\begin{equation}\label{xdot_kin}
     \left(\frac{\dot{x}}{x}\right)^{\text{k}}=\mu_{\text{T}}\cot i\sin(\Theta_{\mu}-\Omega) .
\end{equation}
The constraints introduced by our measurement of $\dot{x}$ are displayed by the solid
orange lines in Figure~\ref{fig:J2045_om_cosi}.

\subsection{Variation of the orbital period}

The observed spin period derivative and orbital period derivative consist of their variations in the
reference frame of the centre of mass of the system (i.e. their intrinsic variations), with additional
contributions from the acceleration due to the transverse velocity of the pulsar (the Shklovskii effect, \citealp{shk70}),
and the difference of the Galactic accelerations of the Solar System and the binary pulsar
projected along the line of sight to the binary pulsar
(e.g.  \citealp{lwj+09}):
\begin{equation} \label{pdot_contributions}
    \left(\frac{\dot{P}_{\text{b}}}{P_{\text{b}}}\right)^{\text{obs}}=\left(\frac{\dot{P_{\text{b}}}}{P_{\text{b}}}\right)^{\text{int}}+\left(\frac{\dot{P_{\text b}}}{P_{\text b}}\right)^{\text{Shk}}+\left(\frac{a}{c}\right)^{\text{Gal}}+\left(\frac{a}{c}\right)^{z} .
\end{equation}
What follows is a brief description of each of the terms in Equation \ref{pdot_contributions}. 
The first term, the intrinsic $\dot{P}_{\text b}$, is made up of several effects, although in this case the dominant effect is the orbital decay caused by the emission of GWs \citep{pm63}:
\begin{equation}\label{pbdot_GW}
    \dot{P}_{\text{b}}^{\text{GW}}=-\frac{192\pi}{5}T_{\odot}^{5/3}\left(\frac{P_{\text{b}}}{2\pi}\right)^{-5/3}f(e)\frac{m_{\text{p}}m_{\text{c}}}{(m_{\text{p}}+m_{\text{c}})^{1/3}},
\end{equation}
where $f(e)$ is the mass function in terms of orbital eccentricity, given by
\begin{equation}\label{massfnecc}
    f(e)=\frac{1+(73/24)e^{2}+(37/96)e^{4}}{(1-e^{2})^{7/2}} .
\end{equation}
This effect mostly becomes significant in very close orbits consisting of compact objects.
For our measured masses of PSR\,J2045+3633 and its companion, and assuming that GR is the correct theory of
gravity, we obtain $\dot{P}_{\text{b}}^{\text{GW}} \, = \, -5.74(15)\times10^{-17}\,\text{s\,s}^{-1}$.
This is negligible compared to our measurement precision.
The same is true for other terms that contribute to $\dot{P}_{\text b}^{\text{int}}$, such as the variation of the orbital period
due to radiative mass loss in the system.

The Shklovskii effect arises from the total proper motion $\mu_T$:
\begin{equation} \label{pdot_shklovskii}
    \left(\frac{\dot{P_{\text b}}}{P_{\text b}}\right)^{\text{Shk}}=\frac{\mu_T^{2} d}{c} .
\end{equation}
From this, we estimate the Shklovskii contribution to $\dot{P}_{\rm b}$
to be $1.37(28)\, \times\, 10^{-13}\,\text{s\,s}^{-1}$.

For a pulsar at a distance $D\equiv R_{0}/\delta$, the component of its acceleration due to the differential disk rotation around the Galactic core is given by \citep{phi92}:
\begin{equation} \label{pdot_galactic}
    \begin{aligned}
    & \left(\frac{a}{c}\right)^{\text{Gal}}=\frac{\boldsymbol{a}_{\text{p}}\cdot\boldsymbol{n}}{c} \\
    & \ \ \ \ \ \ \ \ \ \ \ \ =-A_{\odot}\left[\cos b\cos l+\frac{\delta-\cos b\cos l}{1+\delta^{2}-2\delta\cos b\cos l}\right] ,
    \end{aligned}
\end{equation}
where $R_{0}\, = \, 8.122(31)$\,kpc is the galactocentric distance of the Sun \citep{gra+18}, $l$ and $b$ are the Galactic longitude and latitude respectively, $\boldsymbol{a}_{\text{p}}$ is the pulsar acceleration vector, $\boldsymbol{n}$ is a unit vector pointing from the Solar system barycentre to the pulsar, and
\begin{equation} \label{a_sun}
    A_{\odot}=\frac{V_{\text{c}}^{2}}{cR_{0}} ,
\end{equation}
where, $V_{\text{c}}\, = \, 233.34(14)\,\text{km\,s}^{-1}$ is the Galactic rotation velocity \citep{mcg18}. For this, we estimate a contribution to $\dot{P}_{\text b}$ of $-1.23(15) \, \times \, 10^{-12}\,\text{s\,s}^{-1}$.

The vertical component of the acceleration experienced by a pulsar at a Galactic altitude $z$ is \citep{lwj+09}:
\begin{equation} \label{pdot_zheight}
    \left(\frac{a}{c}\right)^{z} \, = \, - 7.57 \times 10^{-20} |z|  + 12.28 \times 10^{-20} \left( 1 - e^{-4.31  |z|}\right) ,
\end{equation}
which is valid for pulsars at Galactic altitudes $|z| \, \lessapprox \, 1.5$\,kpc \citep{hf04}.
For PSR\,J2045+3633, the Galactic altitude is $-381$\,pc, well within the valid range, and we calculate a contribution to $\dot{P}_{\text b}$ of $-2.43(26)\, \times \, 10^{-14}\,\text{s\,s}^{-1}$.

\begin{figure*}
	\includegraphics[scale=0.68]{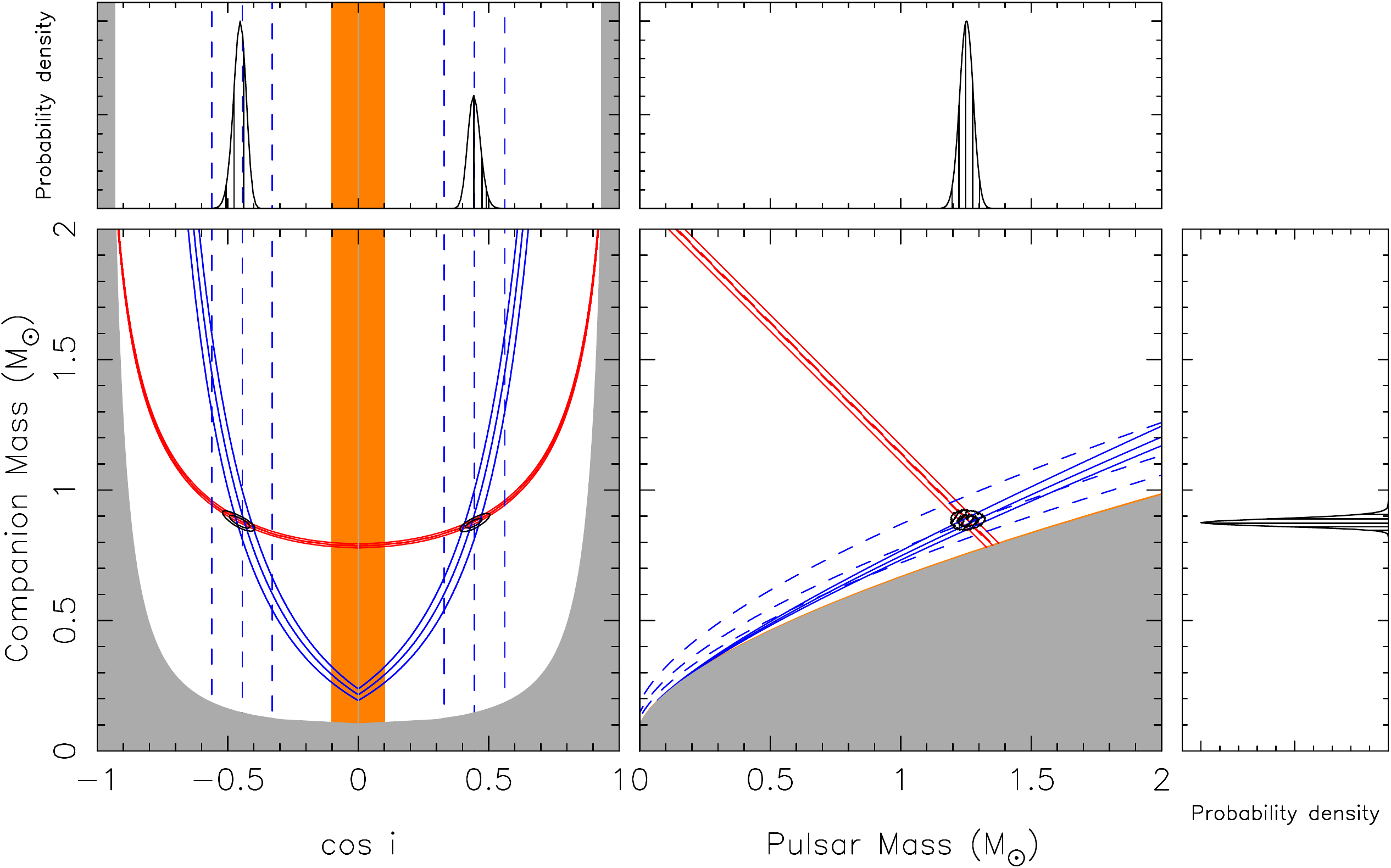}	
	\centering
	\caption{Mass constraints of the PSR\,J2045+3633 companion as a function of the cosine of the orbital inclination (left), and of the PSR\,J2045+3633 mass (right). The grey regions are excluded by the requirement that $m_{\text p}$ is greater than zero (left), and by our measurement of
	the mass function of the system (right), and the orange region is excluded by our $\dot{x}$ measurement. The curves and $1\sigma$ uncertainty regions of the post-Keplerian parameters measured with the DDGR orbital model are overplot; these are $\dot{\omega}$ (solid red), $h_{3}$ (solid blue), and $\varsigma$ (dashed blue). The contours show the 1- and $2\sigma$ likelihood regions, measured from the $\chi^{2}$ mapping with the DDK model, and the marginal probability densities are displayed for each plot axis. The values from both the timing with the DDFWHE model and the $\chi^{2}$ mapping agree very well with each other, with all constraints being consistent with the $1\sigma$ uncertainty regions. The exact orbital inclination is not well-constrained: although the negative value of $\cos i$ is preferred, the polarimetry
	of the pulsar suggests that the positive $\cos i$ (with inclinations around $63^\circ$) is the real value. The pulsar and companion masses are very well constrained, yielding values $m_{\text{p}}=1.251^{+0.021}_{-0.021}$\,M$_{\odot}$ and $m_{\text{c}}=0.873^{+0.016}_{-0.014}$\,M$_{\odot}$. The 1-D probability density function for $\cos i$ is significantly narrower than the uncertainty for $\varsigma$, the reason for that is the additional constraints for the inclination given by the measurement of $\dot{\omega}$ and $h_3$.}
	\label{fig:J2045_om_mass}
\end{figure*}

\begin{figure*}
	\includegraphics[scale=0.9]{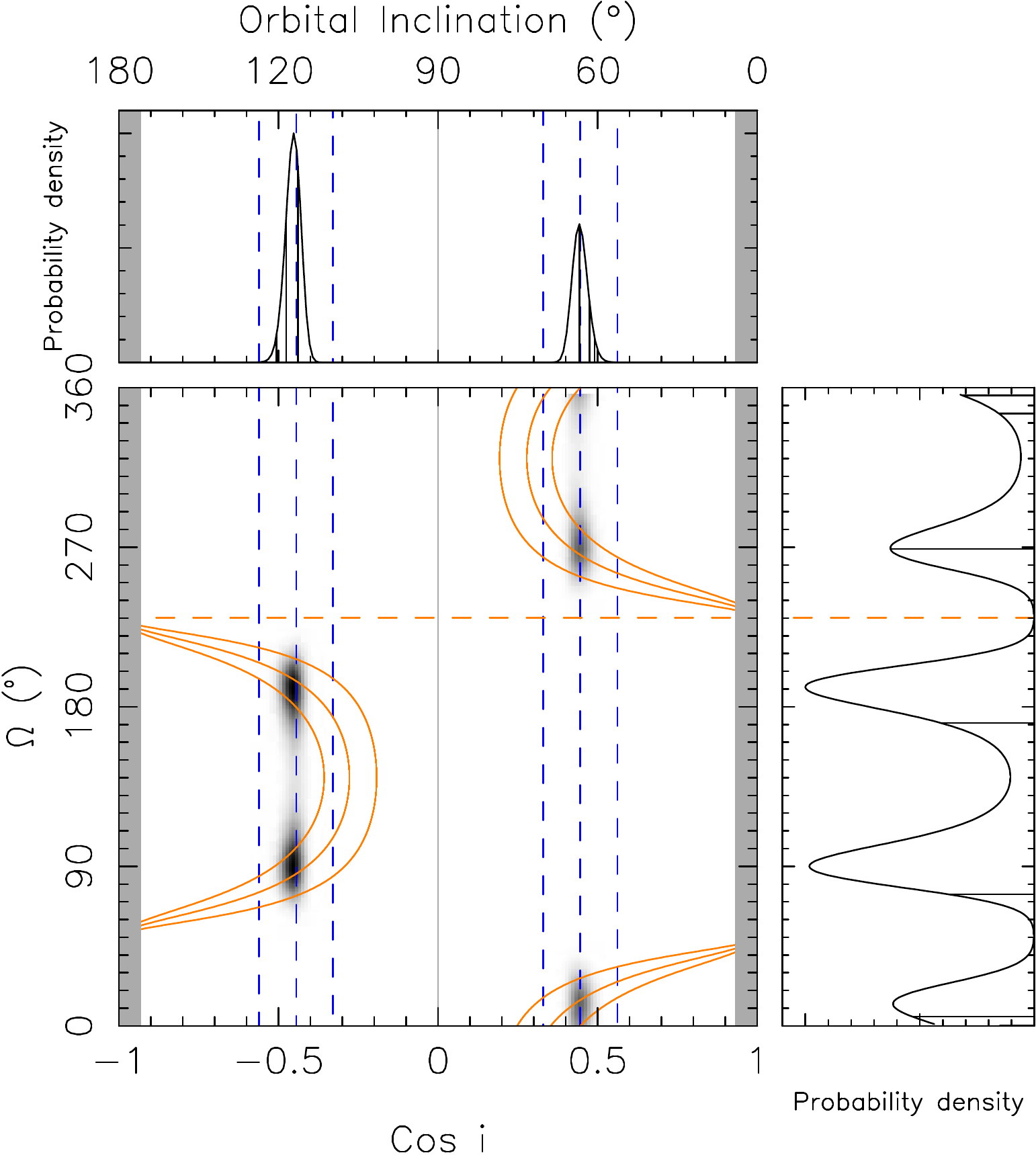}	
	\centering
	\caption{Constraints of the PSR\,J2045+3633 orbital line of nodes, as a function of the cosine of the orbital inclination. The grey shaded region is excluded by the mass function in combination with the estimate for the total mass. The dashed orange line indicates the position angle of the proper motion ($\Theta_{\mu}$), and the curves with $1\sigma$ uncertainty regions are displayed for $\dot{x}$ (solid orange), and $\varsigma$ (dashed blue), obtained from the DDFWHE model. The greyscale colour maps show the likelihood from the $\chi^{2}$ mapping analysis, and are from the same 3-D map shown in Figure \ref{fig:J2045_om_mass}; here with darker shades corresponding to higher probability. The marginal plots show the 1-D probability density functions for both axes. We can see that there are four main areas allowed by our timing analysis, which are well-defined by the intersection of the $\cos i$ and $\dot{x}$ constraints. Currently, we cannot eliminate this degeneracy based on the timing alone, as the system is too distant for the detection of the annual orbital parallax which could break this degeneracy. Although the maximum likelihood occurs at $\Omega\sim198^{\circ}$ and $\cos i\sim-0.45$, the polarimetry implies that the system has a positive $\cos i$, which implies that one of the two positive $\cos i$ solutions
	is the correct one.
	}
	\label{fig:J2045_om_cosi}
\end{figure*}

Putting these together, and taking into account the distance uncertainty, we find that all of the kinematic terms amount to $\dot{P}_{\text b}^{\text{kin}} \, = \, -0.60^{+0.01}_{-0.05} \, \times \, 10^{-12}\,\text{s\,s}^{-1}$. This is
consistent with the results of our timing analysis: with the DDFWHE model, we obtain $\dot{P}_{\text b} \, = \, +3.6(53) \, \times \, 10^{-12}\,\text{s\,s}^{-1}$.
From this, we conclude that we cannot yet detect the $\dot{P}_{\rm b}$ expected from kinematic effects, and that the precision of $\dot{P}_{\text b}$ must be improved by more than an order of magnitude before measurement will be possible. This will be advantageous, as a measurement of the kinematic contribution to $\dot{P_{\text{b}}}$ would enable an accurate measurement of $d$.

A similar calculation is done for the spin period derivative, and the results are presented in Table~\ref{tab:timing_values}.
 
\subsection{The role of spin-orbit coupling}
\label{sec:SO}

In Section \ref{subsec_binary_timing}, we mentioned that for systems consisting of two degenerate stars, we can generally assume that
Newtonian perturbations to a Keplerian orbit can be neglected. However, this is not always the case. For PSR\,J1141$-$6545, a binary pulsar where the
companion is a massive WD \citep{abw+11}, a change in the orbital inclination of the system  was detected,
by assuming the validity of GR for this system, i.e. by using the DDGR model \citep{vbs+20}. Such a change of inclination
is caused (again, assuming the validity of GR) by classical and relativistic spin-orbit effects that result from the fast spin of the companion
WD, which must have a spin period of only a few minutes.  Although this anomalously-fast spin of the companion WD is caused by the
unusual evolutionary history of that system (the WD accreted matter from the progenitor to the pulsar), it demonstrates that
we must always be careful when making this assumption.

Evaluating these spin-orbit contributions for PSR\,J2045+3633 (Equations S7--S9 in \citealt{vbs+20}), we find that, even assuming a spin period of 2 minutes (the lower limit
for the spin period of a spun-up WD like PSR\,J1141$-$6545), the maximum magnitudes
of the classical and
relativistic contributions to $\dot{x}$ are $\sim \,6 \, \times \, 10^{-16} \, \rm lt$-$\rm s \, s^{-1}$
and $\sim \, -7 \, \times \, 10^{-16} \, \rm lt$-$\rm s \, s^{-1}$ respectively. These are an order of magnitude 
smaller than our measured $\dot{x}$. In addition to this, unlike in the case of PSR\,J1141$-$6545,
the WD companion to PSR\,J2045+3633 was not spun up by matter accreted from the progenitor to the pulsar. Instead, it was the pulsar that
accreted matter from the progenitor to the companion WD; we know this because the pulsar shows clear signs of
having been recycled. For this reason, the WD will likely have a much slower
rotation than the companion of PSR\,J1141$-$6545, and therefore the contribution of spin-orbit coupling to the observed $\dot{x}$ and $\dot{\omega}$ can be safely ignored.

\subsection{\text{$\chi^{2}$} mapping} \label{sec:bayesian_mapping}

We can use all the constraints described above to determine the masses and the orbital orientation of the PSR\,J2045+3633 system in a self-consistent
way. In order to do this, we use a $\chi^{2}$ mapping technique similar to the one detailed in \cite{sfa+19}.
In order to take into account the kinematic effects on all parameters (described in \citealp{kop95} and \citealp{kop96}, and which we discuss in Sections \ref{subsec_periastron_advance} and \ref{subsec_semi_major_ax}),
we use the DDK orbital model, but in addition to this, assume the validity of GR.
We map a 3-D parameter space consisting of the total mass of the system ($M$), the cosine of the orbital
inclination ($\cos i$), and the position angle of the line of nodes ($\Omega$). The reason for mapping along the $M$ vector is that
the total mass is well-constrained by the $\dot{\omega}$ measurement of the system (see Section \ref{subsec_periastron_advance}). The reason to map $\Omega$ and $\cos i$
is that this space is, for randomly-aligned orbits, uniformly populated, and thus provides a flat prior.

For each point in this 3-D space, the mass of the pulsar and the companion are known (via Equation~\ref{mass_fn_eqn}), and so we can calculate all PK
parameters according to their GR equations and assume them in a trial DDK model, where they are held constant.
The same is true for all of the kinematic effects (in particular, the secular and annual variations of $x$ and $\omega$), which are
also completely determined for each point in this space. These are calculated within the DDK model from
$\cos i$, $\Omega$, the proper motion, and the assumed parallax (which is the inverse of the DM distance
in kpc), without the need to specify $\dot{x}$.
We then fit this model to the timing data by minimising the residuals; for this, we allow all other timing parameters
(i.e. the astrometric, spin, and Keplerian orbital parameters) to vary. A 3-D probability density is then derived from the
resulting $\chi^2$ map.

Due to the degeneracy between $i$ and $(180^{\circ} - i)$ in the orbital inclination, we have sampled two different $\cos i$ regions: ${[-0.9^{\circ}:-0.1^{\circ}]}$ and ${[+0.1^{\circ}:+0.9^{\circ}]}$, with steps of $0.01^{\circ}$. Within both of these regions, we have sampled the $\Omega$ range ${[0^{\circ}:359^{\circ}]}$ with steps of $1^{\circ}$, and the $M$ range of ${[1.1\,\text{M}_{\odot}:3.3\,\text{M}_{\odot}]}$, with steps of 0.01\,$\text{M}_{\odot}$. The results are summarised in Table \ref{tab:gridresults}.

From our $\chi^{2}$ mapping results (Figure \ref{fig:J2045_om_mass}), we have been able to infer very precise pulsar and companion masses of ${m_{\text{p}} \, = \, 1.251^{+0.021}_{-0.021}\,\text{M}_{\odot}}$ and $m_{\text{c}} \, = \, 0.873^{+0.016}_{-0.014}\,\text{M}_{\odot}$, respectively, and a total system mass of $M \, = \, 2.127^{+0.031}_{-0.031}\,\text{M}_{\odot}$, where all of our quoted uncertainties are $1\sigma$. This represents an order of magnitude improvement in precision over the previously published value \citep{bcf+17}. We are also able to constrain the orbital
inclination $i$ to two relatively narrow intervals: $63.8^{+1.5}_{-1.6}{^{\circ}}$ and $117.5^{+1.6}_{-1.5}{^{\circ}}$, which are much narrower than the estimate of $\varsigma$.
The precise masses and orbital inclination are described very well by the intersection of the $\dot{\omega}$ and $h_3$
constraints (see Figure~\ref{fig:J2045_om_mass}); they cannot be well-described using the $r$-$s$
parameterisation of the Shapiro delay.

Although the two $\cos i$ intervals are relatively narrow, we are not able to well-constrain $\Omega$, and we find four
degenerate ${\Omega\text{-}\cos i}$ solutions, albeit with the values corresponding to a negative $\cos i$ being approximately twice as likely, which can be seen in Figure~\ref{fig:J2045_om_cosi}. The four possible $\Omega$ values are $9^{+13}_{-20}{^{\circ}}$, $92^{+16}_{-13}{^{\circ}}$,  $187^{+13}_{-18}{^{\circ}}$, and $272^{+16}_{-13}{^{\circ}}$. The lower probability of a positive $\cos i$ is not
statistically significant, and therefore our findings are consistent with the orbital inclination derived from the 
polarimetry of the pulsar by \cite{bcf+17} and ourselves (Figure \ref{fig:J2045polprof}, and see Section~\ref{sec:rvm}). Assuming a positive $\cos i$ to therefore be true, the results of the $\chi^{2}$ mapping imply that $i=63.8^{+1.5}_{-1.6} {^{\circ}}$, and that the value of $\Omega$ is either $92^{+16}_{-13}{^{\circ}}$ or  $187^{+13}_{-18}{^{\circ}}$.

Overall, the allowed regions in the $\Omega$-$\cos i$ plane are well-described by the narrow ranges of $\cos i$ determined by the intersection of $\dot{\omega}$ and $h_3$,
and the constraints from the $\dot{x}$ measured in the DDFWHE solution.
In order to determine which of the four combinations of $\Omega$ and $\cos i$ is correct purely
from timing, we would need
to measure the annual orbital parallax of the system (e.g. as \citealt{sfa+19} did for PSR\,J2234+0611).
Given the large distance to PSR\,J2045+3633, the prospects for achieving this are not promising.
It is clear though that continued timing will result in much improved values for $\dot{\omega}$ and $\dot{x}$,
which will further improve the precision of the mass measurements and place tighter constraints on the regions
of the $\Omega$-$\cos i$.

\begin{figure*}
    \advance\leftskip-0.12cm
	\includegraphics[scale=0.25]{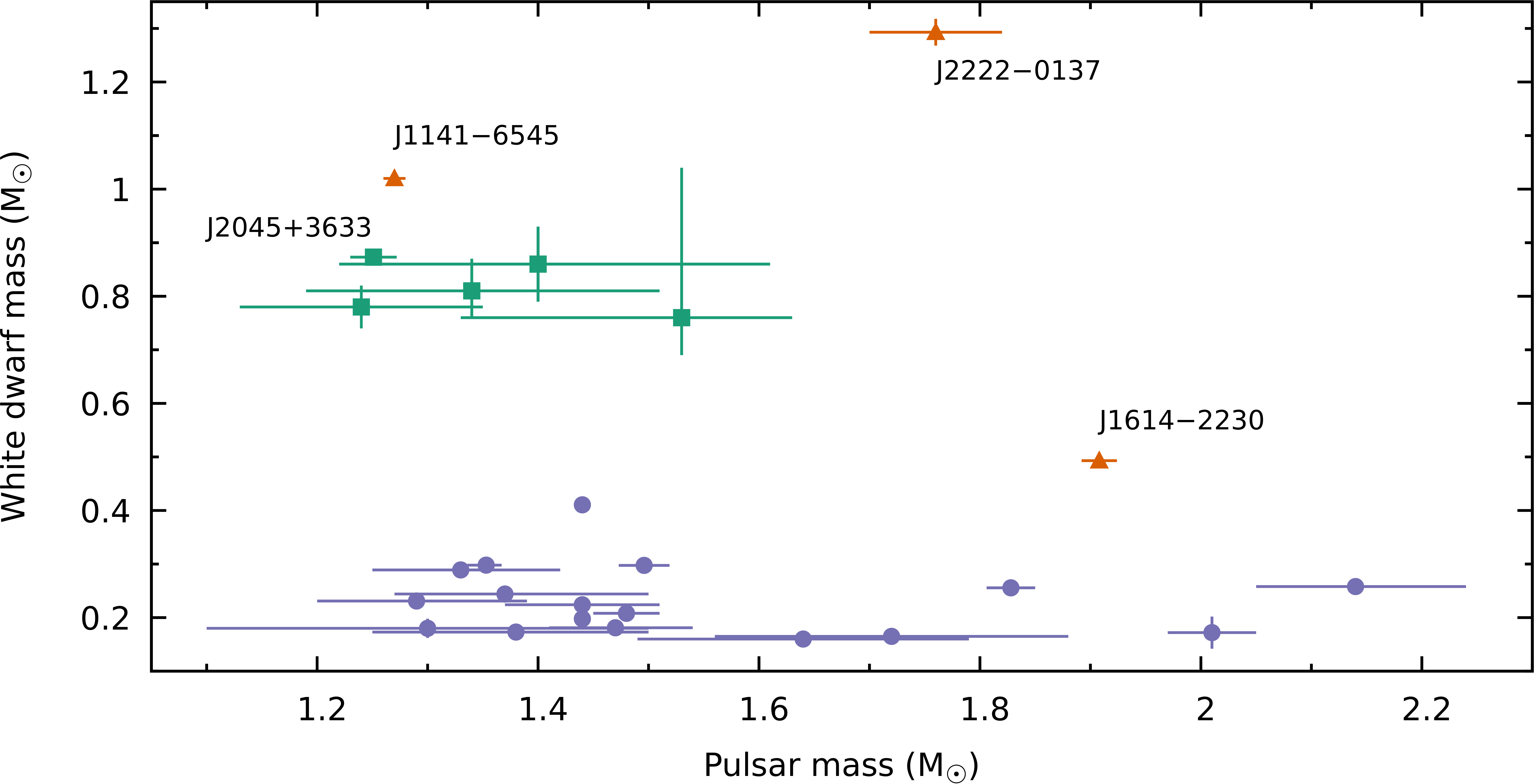}	
	\centering
	\caption{Masses of all known pulsar-white dwarf systems with precise mass measurements that are not located
	in globular clusters. The systems with (low-mass) helium white dwarf (HeWD) companions are shown as purple circles, and the systems with massive white dwarf (CO~WD) companions are shown
	as green squares. With the exception of three special cases, which are highlighted (orange triangles: PSR\,J1141$-$6545, where the white dwarf formed first and thus the pulsar is not recycled, and the unusual systems J1614$-$2230 and J2222$-$0137, discussed in the
	text), all systems fall into two distinct populations. The systems with CO~WDs mostly appear within a narrow range of pulsar and WD masses. }
	\label{fig:psr_wd_masses}
\end{figure*}

\section{Discussion}
\label{sec:discussion}

\subsection{Mass measurements}

One of the main results of this work is the precise mass measurements of PSR\,J2045+3633 and its WD companion. We conclude that, given the $0.873^{+0.016}_{-0.014}\,\text{M}_{\odot}$ mass we have measured for the companion, it is likely to be a carbon-oxygen white dwarf (CO~WD), and not an oxygen-neon-magnesium white dwarf i.e. the progenitor was not sufficiently massive for nuclear carbon burning to be initiated \citep{tlk12}. The $1.251^{+0.021}_{-0.021}\,\text{M}_{\odot}$ mass we have measured for PSR\,J2045+3633 is relatively low, and is evidence for very inefficient accretion of matter during the recycling phase, following its formation. 
Given that NSs are expected to be born with masses in excess of $\sim 1.15-1.20\,\text{M}_\odot$ \citep[][and references therein]{tj19}, this indicates that the pulsar accreted less than $\sim0.10\,\text{M}_\odot$, despite significant mass loss from the progenitor star of the $0.873\,\text{M}_\odot$ CO~WD, which most likely had a ZAMS mass in excess of $4\,\text{M}_\odot$ (see e.g. Figure~1 in \citealp{tlk11}).

To further put the masses of PSR\,J2045+3633 and its companion into perspective, in Figure~\ref{fig:psr_wd_masses} we plot all currently-known
precise masses measured for pulsar-WD systems.
In this plot, we can see several striking trends. First, most systems fall into two
well-defined populations. The first corresponds to systems with Helium WD (HeWD) components, which have a narrow distribution of
WD masses (which are mostly correlated with the orbital period of the system, as described by \citealp{ts99}). The second
population has a surprisingly narrow range of companion CO~WD masses, from 0.7 to 0.9\,$\text{M}_{\odot}$, and
these masses are not correlated with the orbital period. PSR\,J2043+3633
is clearly within this latter class, and this work provides the best-measured masses of systems that are within this group.
Another member of this class with a well-known mass, PSR\,J2053+4650,
was studied in detail in the same work that announced the discovery of
PSR\,J2045+3633 \citep{bcf+17}.

There is a relative dearth of WD masses between 0.41 and 0.7\,$\text{M}_{\odot}$. This gap is
partly observational: according to the \cite{ts99} relation between the orbital period of
a pulsar-HeWD binary and the HeWD mass, we expect the upper end of this range (up to $0.46\,\text{M}_\odot$) to be populated exclusively by pulsar-HeWD binaries in orbits with periods in excess of a few hundreds of days.
The observed distribution of mass functions suggests that this is indeed the
case. However, the lack of precisely-measured WD and pulsar masses with such wide orbits is caused by the fact that, for those wide orbits, pulsar recycling is inefficient and produces pulsars with relatively short-lived radio emission (i.e. due to insufficient decay of the remaining magnetic field, the spin-down torque is large), and with spin periods of many tens of milliseconds. For those slow-spinning pulsars, the achievable timing precision
is generally not sufficiently high for the precise detection of the Shapiro delay signature.
Furthermore, with large orbital periods, there are potentially large variations in DM during the
long orbital cycle, which further complicates Shapiro delay measurements.

A slow spin period is also an issue for pulsars with massive WD companions, since the massive progenitor to the white dwarf evolves rapidly and the recycling process is relatively short. However, as is the case for PSR\,J2045+3633, some of these systems have a
measurable $\dot{\omega}$ which, together with low-precision detections of the Shapiro delay,
allows precise masses to be determined. The $\dot{\omega}$ cannot be measured
for pulsar-HeWDs with orbits of many hundreds of days.

The range of pulsar masses with HeWD companions appears to be distinctively larger than that of pulsars with more
massive CO~WD companions between 0.7 - $0.9\,\text{M}_{\odot}$. It is interesting to speculate on possible causes for this. One
possibility is that the progenitors of HeWDs have a much slower evolution; in these cases, the
accretion episode that is observable as a low-mass X-ray binary phase is very long-lived. This results in much
faster spin periods for the pulsars in this class compared to the pulsars with more massive companions \citep[see e.g. Figures 3 and 9 in][]{tlk12}, and
this could also result in much larger accreted masses. 

Mass transfer between two stars via Roche lobe overflow (RLO) can be initiated while the donor star is still on the main sequence (Case~A RLO), during hydrogen shell burning (Case~B RLO) or during helium shell burning (Case~C RLO). The corresponding evolutionary timescales for these different cases will in general proceed on a nuclear, thermal, or dynamical timescale, respectively, or a combination thereof. 
The progenitor systems of pulsar-CO~WD binaries are thought to be intermediate-mass X-ray binaries (IMXBs), with typical donor star masses of 3 to $6\,\text{M}_{\odot}$. Only a small fraction of these will undergo RLO Case A (during hydrogen core burning), as few such binaries are formed in the narrow orbital range that is required for this scenario. A rare example of this scenario being met is PSR\,J1614$-$2230 \citep{tlk11}, which resulted in the formation of a $0.493(3) \, \rm M_{\odot}$ CO~WD, the only in the gap between 0.41 and $0.7\,\text{M}_{\odot}$ . This is a massive pulsar
\citep{dpr+10,abb+18} that has an unusually fast spin for a system with such a massive WD
companion. The prolonged mass transfer associated with Case A RLO provides a good explanation of these unusual characteristics.

Case B (and C) RLO covers a very wide range of orbital periods, such that the donor star cores can grow to a larger mass, consistent with the observed 0.7 to $0.9\,\text{M}_{\odot}$ range (see Figure 1 of \citealp{tlk11}).
Combined with the initial mass function, it is perhaps reasonable that few very massive WDs are expected, which is consistent with the observed population.

In addition to PSR~J1614$-$2230, there are two unusual pulsar-WD binary systems that do not fall in these two populations (highlighted with orange triangles in Figure~\ref{fig:psr_wd_masses}).
The first one, PSR\,J1141$-$6545,
had a rather unusual evolution, where the
WD formed first and accreted matter from the progenitor of the pulsar, spinning
up significantly in the process \citep{vbs+20}. After that, the secondary exploded in a supernova, forming a
normal pulsar which can not be recycled and therefore remains in an eccentric orbit.
Currently, there are at most two other systems that are thought to have followed a similar evolution
(see \citealt{nct+18} and references therein).
The second unusual system is J2222$-$0137, which has an unusually massive ($\sim \, 1.3 \, \rm  M_{\odot}$)
WD companion, and is the most massive double-degenerate binary system currently known (3.06\,M$_{\odot}$, \citealp{cfg+17}).

To summarise, the observed masses of components in pulsar-WD binaries are consistent with there being two main groups: the pulsar-HeWD systems, where the companion mass is determined by the orbital period of the system and the
range of pulsar masses is quite broad, and a second group of systems with massive WDs that has
a rather narrow range of pulsar and WD masses, with the latter all falling within 0.7 to 0.9\,M$_{\odot}$.
In addition to these two main groups, there are likely several other groups with different evolutionary
histories and (presumably) much smaller populations. A possible exception is the group
of systems with very high-mass WDs, such as PSR\,J2222$-$0137. This is the sole member of that class with a
well-measured mass. However, it is located only 267\,pc from our Solar System, and this proximity implies that there are
probably many more systems like this in the Galaxy i.e. that this class is potentially much more numerous.

\subsection{Transverse velocity}

The PSR\,J2045+3633 transverse velocity of 109(22)\,km\,s$^{-1}$ is calculated in a heliocentric reference frame. In order to derive
further conclusions, we first need to estimate its velocity relative to the local standard of rest (LSR).
To do this, we follow the procedure highlighted by \cite{zfk+19}. 

For an object at the position of the
pulsar that is in the LSR, we expect a horizontal proper motion of $-$5.39\,mas\,yr$^{-1}$, and a vertical proper motion of $-$0.28\,mas\,yr$^{-1}$.
Subtracting these values from the horizontal and vertical components of the observed proper motion, we 
obtain the peculiar proper motion: $+1.30$ and $+1.09$\,mas\,yr$^{-1}$ along the horizontal and vertical
directions respectively i.e. the total peculiar proper motion is around 1.70\,mas\,yr$^{-1}$. This corresponds to 
a transverse peculiar velocity of 44(9)\,km\,s$^{-1}$, which is relatively small in comparison
to most other pulsars, in particular binary pulsars. 
There are other members of the pulsar-massive WD class that are known to have very low
peculiar velocities. Two examples are PSR\,J1802$-$2124 \citep{fsk+10} and
PSR\,J1949+3106; the latter has a tangential peculiar velocity smaller
than 10\,km\,s$^{-1}$ \citep{zfk+19}. More broadly, the pulsars with massive WDs seem to have a smaller
range of Galactic heights, which is consistent with having lower peculiar velocities \citep{nbb+14,pkr+19}.
This can generally be interpreted as the result of the ``anchoring'' effect of a more massive companion at the time
of the supernova that forms the pulsar i.e. the more massive the companion, the smaller the
change of velocity for the system after the supernova (from conservation of momentum). 
An additional effect would be if the progenitors of these more massive WDs were more efficient in
stripping off the envelope of the progenitors to the pulsars, as such stripped supernovae are
likely to have experienced smaller associated kicks \citep[see discussions in][and references therein]{tlp+15}.

\subsection{Binary evolution} \label{subsec_binary_evol}

As the PSR\,J2045+3633 orbit is relatively circular and the pulsar is recycled, albeit only partially ($P=31.7$\,ms, $\dot{P}=5.8847(3)\time10^{-19}\,\text{s\,s}^{-1}$), this indicates that mass transfer occurred after formation of the pulsar, and therefore the companion formed second. 
The large mass of the WD companion implies that it would have formed from an asymptotic giant branch (AGB) star. 

Pulsar-CO~WD binaries are usually thought to form via two evolutionary channels \citep{tlk12}: stable RLO in an IMXB, and common envelope (CE) evolution. Recent work by \cite{mft+20} on stable RLO from IMXB systems demonstrates that it is difficult to produce a binary with the component masses and orbital period that we measure, which makes formation via CE evolution the more likely scenario.

The radius of the progenitor star can be estimated from models relating the core mass to the radius. Using the models presented in \cite{vvp06}, we estimate the stellar radius of the progenitor to be $\sim400\,R_{\odot}=928$\,lt-s 
i.e. approximately an order of magnitude greater than the current size of the orbit ($55\,R_{\odot}=127.6$\,lt-s, calculated from Kepler's third law in Section \ref{subsec_periastron_advance}). Therefore after the companion star evolved to the AGB phase, the system must have gone through a CE phase, during which in-spiralling occurred (e.g. \citealp{pac76}, \citealp{il93}); a scenario which is consistent with the pulsar being only partially recycled. We note, however, that the \cite{vvp06} models are not directly applicable to this system, and we use them here only to obtain a general estimate of the progenitor radius.

For such a formation scenario it is perhaps unexpected that the orbit is still as wide as it is, and with a relatively large eccentricity. Binary systems that are thought to form following a CE phase typically have orbital periods $\sim1$\,--\,10\,days, and much lower eccentricities (see e.g. \citealp{tkr00}). Logically, there are three possible explanations for this:
\begin{enumerate}
	\item The system was initially highly eccentric, and was later stabilised
	at some point in its evolution.
	\item The system was initially more circular, and eccentricity was induced at some point in the binary evolution.
	\item The system was initially moderately eccentric, and much of that eccentricity remains.

\end{enumerate}
Below, we suggest possible scenarios that could arise from points (i)-(iii):

\subsubsection{Stabilised orbit: equilibrium from fluctuation-dissipation}
Low-mass binary pulsars (i.e. those with HeWD companions) typically have a small but non-zero eccentricity, as the envelope of the red-giant branch star progenitor of the companion is convective, and therefore has a non-zero and variable quadrupole moment \citep{phi92}. A statistical relation was found to exist between eccentricity and orbital period for these systems \citep{pk94}, which is invalid for binaries with higher-mass companions (e.g. CO~WDs), which will have a different structure at the point where they detach. While the \cite{phi92} process does not apply to binaries with high-mass WD companions, these systems do appear to have a preference for higher eccentricities and longer periods in higher-mass systems, which could indicate that a similar process occurs during their evolution.

\subsubsection{Induced eccentricity}
It is possible that the orbit was circularised in the  post-CE phase, but some process induced eccentricity later in the evolution of the system. This could perhaps be somewhat similar to the process described by \cite{ant14} for eccentric pulsar-HeWD binaries, where late in the evolution, a large final flash expels material from a proto-HeWD companion into a circumbinary disk, which induces eccentricity. Though this model cannot work directly for PSR\,J2045+3633, since here a helium shell flash would be needed when producing a CO~WD, and it remains to be shown if such a scenario is possible.

Related to this scenario, eccentricity could have been induced in the post-CE phase through interaction with a post-AGB phase circumbinary dust ring remnant, through the process described by \cite{dij+13}. This scenario would imply that the CE triggered by AGB phase was the last stage of mass transfer in the evolution of the system (i.e. there was no naked helium giant phase, so-called Case~BB RLO).

\subsubsection{Leftover eccentricity: rapid common envelope phase}
A short CE phase in the binary evolution would not allow enough time for the binary to be circularised through friction in the envelope, and would account for the pulsar being only partially recycled (see \citealp{tlk12} for a similar hypothesis for PSR~J1822$-$0848).
We now investigate this hypothesis as a proof of concept by using simple arguments and our measurements of the pulsar and WD masses, and the orbital separation.

Earlier, we estimated the stellar radius of the AGB star to be $\sim 400$\,R$_{\odot}=928$\,lt-s, using the models presented in \cite{vvp06}. We also use the measured WD mass to estimate a progenitor AGB star mass of roughly  
$\sim 4.9\,\text{M}_\odot$
using the models presented in \cite{klp02}. If we assume that the Roche-lobe radius is equal to the radius of the AGB star, this allows us to estimate the initial separation $a_{\text{i}}$ of the binary components using \citep{egg83}
\begin{equation}\label{eggleton_eqn}
    R_{\text{rl}}=\frac{0.49q^{2/3}}{0.6q^{2/3}+\ln{(1+q^{1/3})}}\;a_{\text{i}} ,
\end{equation}
where $q=M_{1}/M_{2}$ is the mass ratio (donor star mass to accretor star mass), i.e. we use $q=4.9/1.25$.
Using this, we estimate an initial component separation $a_{\text{i}}=800\,\text{R}_{\odot}=1857$\,lt-s.

We can test this scenario by comparing the energy required to unbind the CE to the initial and final orbital binding energies. The change in orbital energies is given by
\begin{equation}\label{changeinenergy}
    \Delta E_{\text{orbit}}=\frac{GM_{\text{wd}}M_{\text{psr}}}{2a_{\text{f}}}-\frac{GM_{\text{agb}}M_{\text{psr}}}{2a_{\text{i}}} ,
\end{equation}
and taking the final orbital separation as our measured value of the PSR\,J2045+3633 orbital semi-major axis, $a_{\text{f}}=55\,R_{\odot}=127.6$\,lt-s, we estimate a change in orbital energy $\Delta E=2.31\times10^{46}$\,erg. The CE binding energy is given by \citep{dek90}
\begin{equation}\label{bindingenergy}
    E_{\text{bind}}=\frac{GM_{\text{agb}}M_{\text{wd}}}{\lambda R_{\text{rl}}} ,
\end{equation}
where $\lambda$ is a numerical value that is often (but not always) of order unity, which accounts for the structure of the star, and is usually either left as a free parameter or estimated through detailed modelling (see e.g. \citealp{dt00,ijc+13}). Assuming $\lambda=1$ and using our earlier values, we estimate a value for the CE binding energy $E_{\text{bind}}=2.02\times10^{46}$\,erg. This is remarkably close to our estimate of the change in orbital energy during the in-spiral, and the two values can be made to fit by setting $\lambda=0.87$, and assuming 100\% efficiency in the conversion of released orbital energy into kinetic energy to expel the CE.

Although merely based on simplified energy considerations and disregarding additional effects such as released recombination energy, accretion energy and, not least, a detailed donor star structure (see \citealp{ijc+13} for a review), we conclude that this CE scenario could be possible. 
The expected outcome would indeed be a large reduction in orbital separation, and a mildly-recycled pulsar in a non-zero eccentric orbit due to rapid in-spiral.

We note that a more realistic model of the $\lambda$ parameter for evolved (giant) stars predicts values much larger than unity ($\lambda\sim10$ for $4-5\,\text{M}_\odot$ stars at $R\sim400\,\text{R}_\odot$, \citealp{dt00}), which would significantly lower the binding energy of the CE and facilitate its ejection and the proposed scenario.
A thorough evaluation of the orbital evolution is beyond the scope of our work.

\section{Conclusions and prospects}
\label{sec:conclusions}

Using 6 years of radio timing data, we have made precise mass measurements of PSR\,J2045+3633 and its WD companion, and have placed further constraints on its orbital geometry. The data set we have used in this work has allowed an order of magnitude improvement over the previously-published masses.

We have examined the mass-mass distribution of Galactic pulsar-WD systems, and find evidence for most systems with precisely-measured masses belonging to two distinct populations; one of them, the class with CO WD companions to which PSR~J2045+3633 belongs, with a surprising narrow range of WD masses.
Using our measurements of the masses and orbital parameters, we have proposed scenarios for the evolutionary history of this binary system, and conclude that a rapid common envelope phase is the most likely formation scenario.

Continued timing will greatly improve the measurements of the PK parameters, particularly $\dot{\omega}$ and $\dot{x}$.
This will yield much improved mass estimates for this system, and significantly reduce the
sizes of the allowed regions of the $\Omega$--$\cos i$ plane (Figures \ref{fig:J2045_om_mass} and \ref{fig:J2045_om_cosi}), but it is not clear whether this will allow the degeneracy to be eliminated
between the four $\Omega$ and $\cos i$ solutions, based on timing alone. Two of those solutions can, in principle,
be eliminated based on the results of the rotating vector model fit to the polarimetric data of the pulsar.

As this pulsar is distant, we have not been able to measure a parallax from the timing,
meaning that many of the derived quantities which rely on distance are only estimates.
However, the continued and relatively fast improvement of 
$\dot{P_{\text{b}}}$ will eventually yield a precise distance (e.g. \citealp{bb96}).
Together with the improvement in the measurement of the proper motion, this will yield much improved
estimates of the velocity of the system. All of these will be useful to improve our models
of the evolution of the binary.

\section*{Acknowledgements}
We thank Marten van Kerkwijk for valuable discussions.
The Arecibo Observatory is operated by the University of Central Florida, Ana G. M\`{e}ndez-Universidad Metropolitana, and Yang Enterprises under a cooperative agreement with the National Science Foundation (NSF; AST-1744119). We especially thank the dedicated and professional on-site staff, who have overcome many challenges and difficult circumstances in order to enable the continuation of top-quality science with this telescope.
Pulsar research at the Jodrell Bank Centre for Astrophysics is supported by a consolidated grant from STFC.
The Nan\c{c}ay Radio Observatory is operated by the Paris Observatory, associated with the French Centre National de la Recherche Scientifique (CNRS). We acknowledge financial support from the ``Programme National Gravitation, R\'ef\'erences, Astronomie, M\'etrologie'' (PNGRAM) of CNRS/INSU, France. 
Part of the data taken for this work is based on observations with the 100-m telescope of the MPIfR (Max-Planck-Institut f{\"u}r Radioastronomie) at Effelsberg. We thank the anonymous referee for their very helpful comments which have improved how our work is presented here.

\section*{Data Availability}
The data underlying this article will be shared on reasonable request to the corresponding author.

\bibliographystyle{mnras}
\bibliography{psrrefs}{}

\begin{thebibliography}{}
\makeatletter
\relax
\def\mn@urlcharsother{\let\do\@makeother \do\$\do\&\do\#\do\^\do\_\do\%\do\~}
\def\mn@doi{\begingroup\mn@urlcharsother \@ifnextchar [ {\mn@doi@}
  {\mn@doi@[]}}
\def\mn@doi@[#1]#2{\def\@tempa{#1}\ifx\@tempa\@empty \href
  {http://dx.doi.org/#2} {doi:#2}\else \href {http://dx.doi.org/#2} {#1}\fi
  \endgroup}
\def\mn@eprint#1#2{\mn@eprint@#1:#2::\@nil}
\def\mn@eprint@arXiv#1{\href {http://arxiv.org/abs/#1} {{\tt arXiv:#1}}}
\def\mn@eprint@dblp#1{\href {http://dblp.uni-trier.de/rec/bibtex/#1.xml}
  {dblp:#1}}
\def\mn@eprint@#1:#2:#3:#4\@nil{\def\@tempa {#1}\def\@tempb {#2}\def\@tempc
  {#3}\ifx \@tempc \@empty \let \@tempc \@tempb \let \@tempb \@tempa \fi \ifx
  \@tempb \@empty \def\@tempb {arXiv}\fi \@ifundefined
  {mn@eprint@\@tempb}{\@tempb:\@tempc}{\expandafter \expandafter \csname
  mn@eprint@\@tempb\endcsname \expandafter{\@tempc}}}

\bibitem[\protect\citeauthoryear{{Alam} et~al.,}{{Alam} et~al.}{2020}]{aab+20}
{Alam} M.~F.,  et~al., 2020, arXiv e-prints, \href
  {https://ui.adsabs.harvard.edu/abs/2020arXiv200506495A} {p. arXiv:2005.06495}

\bibitem[\protect\citeauthoryear{{Alpar}, {Cheng}, {Ruderman}  \&
  {Shaham}}{{Alpar} et~al.}{1982}]{alp82}
{Alpar} M.~A.,  {Cheng} A.~F.,  {Ruderman} M.~A.,   {Shaham} J.,  1982, \mn@doi
  [\nat] {10.1038/300728a0}, \href
  {https://ui.adsabs.harvard.edu/abs/1982Natur.300..728A} {300, 728}

\bibitem[\protect\citeauthoryear{{Anderson}, {Freire}  \& {Yunes}}{{Anderson}
  et~al.}{2019}]{afy19}
{Anderson} D.,  {Freire} P.,   {Yunes} N.,  2019, \mn@doi [Classical and
  Quantum Gravity] {10.1088/1361-6382/ab3a1c}, \href
  {https://ui.adsabs.harvard.edu/abs/2019CQGra..36v5009A} {36, 225009}

\bibitem[\protect\citeauthoryear{{Antoniadis}}{{Antoniadis}}{2014}]{ant14}
{Antoniadis} J.,  2014, \mn@doi [\apjl] {10.1088/2041-8205/797/2/L24}, \href
  {https://ui.adsabs.harvard.edu/abs/2014ApJ...797L..24A} {797, L24}

\bibitem[\protect\citeauthoryear{{Antoniadis}, {Bassa}, {Wex}, {Kramer}  \&
  {Napiwotzki}}{{Antoniadis} et~al.}{2011}]{abw+11}
{Antoniadis} J.,  {Bassa} C.~G.,  {Wex} N.,  {Kramer} M.,   {Napiwotzki} R.,
  2011, \mn@doi [\mnras] {10.1111/j.1365-2966.2010.17929.x}, \href
  {https://ui.adsabs.harvard.edu/abs/2011MNRAS.412..580A} {412, 580}

\bibitem[\protect\citeauthoryear{{Antoniadis} et~al.,}{{Antoniadis}
  et~al.}{2013}]{afw+13}
{Antoniadis} J.,  et~al., 2013, \mn@doi [Science] {10.1126/science.1233232},
  \href {https://ui.adsabs.harvard.edu/abs/2013Sci...340..448A} {340, 448}

\bibitem[\protect\citeauthoryear{{Antoniadis}, {Tauris}, {Ozel}, {Barr},
  {Champion}  \& {Freire}}{{Antoniadis} et~al.}{2016}]{ato+16}
{Antoniadis} J.,  {Tauris} T.~M.,  {Ozel} F.,  {Barr} E.,  {Champion} D.~J.,
  {Freire} P.~C.~C.,  2016, preprint, \href
  {http://adsabs.harvard.edu/abs/2016arXiv160501665A} {} (\mn@eprint {arXiv}
  {1605.01665})

\bibitem[\protect\citeauthoryear{{Archibald} et~al.,}{{Archibald}
  et~al.}{2018}]{agh+18}
{Archibald} A.~M.,  et~al., 2018, \mn@doi [\nat] {10.1038/s41586-018-0265-1},
  \href {https://ui.adsabs.harvard.edu/abs/2018Natur.559...73A} {559, 73}

\bibitem[\protect\citeauthoryear{{Arzoumanian} et~al.,}{{Arzoumanian}
  et~al.}{2018}]{abb+18}
{Arzoumanian} Z.,  et~al., 2018, \mn@doi [\apjs] {10.3847/1538-4365/aab5b0},
  \href {https://ui.adsabs.harvard.edu/abs/2018ApJS..235...37A} {235, 37}

\bibitem[\protect\citeauthoryear{{Barr} et~al.,}{{Barr} et~al.}{2013}]{bck+13}
{Barr} E.~D.,  et~al., 2013, \mn@doi [\mnras] {10.1093/mnras/stt1440}, \href
  {https://ui.adsabs.harvard.edu/abs/2013MNRAS.435.2234B} {435, 2234}

\bibitem[\protect\citeauthoryear{{Bassa} et~al.,}{{Bassa}
  et~al.}{2016}]{bjk+16}
{Bassa} C.~G.,  et~al., 2016, \mn@doi [MNRAS] {10.1093/mnras/stv2755}, \href
  {http://adsabs.harvard.edu/abs/2016MNRAS.456.2196B} {456, 2196}

\bibitem[\protect\citeauthoryear{{Bell} \& {Bailes}}{{Bell} \&
  {Bailes}}{1996}]{bb96}
{Bell} J.~F.,  {Bailes} M.,  1996, \mn@doi [\apjl] {10.1086/309862}, \href
  {https://ui.adsabs.harvard.edu/abs/1996ApJ...456L..33B} {456, L33}

\bibitem[\protect\citeauthoryear{{Berezina} et~al.,}{{Berezina}
  et~al.}{2017}]{bcf+17}
{Berezina} M.,  et~al., 2017, \mn@doi [\mnras] {10.1093/mnras/stx1518}, \href
  {https://ui.adsabs.harvard.edu/abs/2017MNRAS.470.4421B} {470, 4421}

\bibitem[\protect\citeauthoryear{{Cognard} et~al.,}{{Cognard}
  et~al.}{2017}]{cfg+17}
{Cognard} I.,  et~al., 2017, \mn@doi [\apj] {10.3847/1538-4357/aa7bee}, \href
  {https://ui.adsabs.harvard.edu/abs/2017ApJ...844..128C} {844, 128}

\bibitem[\protect\citeauthoryear{{Cordes} \& {Lazio}}{{Cordes} \&
  {Lazio}}{2002}]{cl02}
{Cordes} J.~M.,  {Lazio} T.~J.~W.,  2002, ArXiv Astrophysics e-prints, \href
  {http://adsabs.harvard.edu/abs/2002astro.ph..7156C} {}

\bibitem[\protect\citeauthoryear{{Cromartie} et~al.,}{{Cromartie}
  et~al.}{2020}]{cfr+20}
{Cromartie} H.~T.,  et~al., 2020, \mn@doi [Nature Astronomy]
  {10.1038/s41550-019-0880-2}, \href
  {https://ui.adsabs.harvard.edu/abs/2020NatAs...4...72C} {4, 72}

\bibitem[\protect\citeauthoryear{{Damour}}{{Damour}}{2015}]{dam15}
{Damour} T.,  2015, \mn@doi [Classical and Quantum Gravity]
  {10.1088/0264-9381/32/12/124009}, \href
  {https://ui.adsabs.harvard.edu/abs/2015CQGra..32l4009D} {32, 124009}

\bibitem[\protect\citeauthoryear{Damour \& Deruelle}{Damour \&
  Deruelle}{1985}]{dd85}
Damour T.,  Deruelle N.,  1985, Ann. Inst. H. Poincar\'e (Physique
  Th\'eorique), 43, 107

\bibitem[\protect\citeauthoryear{Damour \& Deruelle}{Damour \&
  Deruelle}{1986}]{dd86}
Damour T.,  Deruelle N.,  1986, Ann. Inst. H. Poincar\'e (Physique
  Th\'eorique), 44, 263

\bibitem[\protect\citeauthoryear{{Damour} \& {Taylor}}{{Damour} \&
  {Taylor}}{1992}]{dt92}
{Damour} T.,  {Taylor} J.~H.,  1992, \mn@doi [\prd] {10.1103/PhysRevD.45.1840},
  \href {https://ui.adsabs.harvard.edu/abs/1992PhRvD..45.1840D} {45, 1840}

\bibitem[\protect\citeauthoryear{{Demorest}, {Pennucci}, {Ransom}, {Roberts}
  \& {Hessels}}{{Demorest} et~al.}{2010}]{dpr+10}
{Demorest} P.~B.,  {Pennucci} T.,  {Ransom} S.~M.,  {Roberts} M.~S.~E.,
  {Hessels} J.~W.~T.,  2010, \mn@doi [\nat] {10.1038/nature09466}, \href
  {https://ui.adsabs.harvard.edu/abs/2010Natur.467.1081D} {467, 1081}

\bibitem[\protect\citeauthoryear{{Dermine}, {Izzard}, {Jorissen}  \& {Van
  Winckel}}{{Dermine} et~al.}{2013}]{dij+13}
{Dermine} T.,  {Izzard} R.~G.,  {Jorissen} A.,   {Van Winckel} H.,  2013,
  \mn@doi [\aap] {10.1051/0004-6361/201219430}, \href
  {https://ui.adsabs.harvard.edu/abs/2013A&A...551A..50D} {551, A50}

\bibitem[\protect\citeauthoryear{{Desvignes} et~al.,}{{Desvignes}
  et~al.}{2016}]{dcl+16}
{Desvignes} G.,  et~al., 2016, \mn@doi [MNRAS] {10.1093/mnras/stw483}, \href
  {http://adsabs.harvard.edu/abs/2016MNRAS.458.3341D} {458, 3341}

\bibitem[\protect\citeauthoryear{{Dewi} \& {Tauris}}{{Dewi} \&
  {Tauris}}{2000}]{dt00}
{Dewi} J. D.~M.,  {Tauris} T.~M.,  2000, 360, 1043

\bibitem[\protect\citeauthoryear{{Donner} et~al.,}{{Donner}
  et~al.}{2019}]{dvt+19}
{Donner} J.~Y.,  et~al., 2019, \mn@doi [\aap] {10.1051/0004-6361/201834059},
  \href {https://ui.adsabs.harvard.edu/abs/2019A&A...624A..22D} {624, A22}

\bibitem[\protect\citeauthoryear{Eggleton}{Eggleton}{1983}]{egg83}
Eggleton P.~P.,  1983, ApJ, 268, 368

\bibitem[\protect\citeauthoryear{{Ferdman} et~al.,}{{Ferdman}
  et~al.}{2010}]{fsk+10}
{Ferdman} R.~D.,  et~al., 2010, \mn@doi [\apj] {10.1088/0004-637X/711/2/764},
  \href {https://ui.adsabs.harvard.edu/abs/2010ApJ...711..764F} {711, 764}

\bibitem[\protect\citeauthoryear{{Folkner}, {Jacobson}, {Park}  \&
  {Williams}}{{Folkner} et~al.}{2016}]{fjp+16}
{Folkner} W.,  {Jacobson} R.~A.,  {Park} R.,   {Williams} J.~G.,  2016, in
  AAS/Division for Planetary Sciences Meeting Abstracts \#48. AAS/Division for
  Planetary Sciences Meeting Abstracts.
p. 120.07

\bibitem[\protect\citeauthoryear{{Fonseca}, {Stairs}  \& {Thorsett}}{{Fonseca}
  et~al.}{2014}]{fst14}
{Fonseca} E.,  {Stairs} I.~H.,   {Thorsett} S.~E.,  2014, \mn@doi [\apj]
  {10.1088/0004-637X/787/1/82}, \href
  {https://ui.adsabs.harvard.edu/abs/2014ApJ...787...82F} {787, 82}

\bibitem[\protect\citeauthoryear{{Freire} \& {Wex}}{{Freire} \&
  {Wex}}{2010}]{fw10}
{Freire} P. C.~C.,  {Wex} N.,  2010, \mn@doi [\mnras]
  {10.1111/j.1365-2966.2010.17319.x}, \href
  {https://ui.adsabs.harvard.edu/abs/2010MNRAS.409..199F} {409, 199}

\bibitem[\protect\citeauthoryear{{Freire} et~al.,}{{Freire}
  et~al.}{2011}]{fbw+11}
{Freire} P.~C.~C.,  et~al., 2011, \mn@doi [\mnras]
  {10.1111/j.1365-2966.2010.18109.x}, \href
  {https://ui.adsabs.harvard.edu/abs/2011MNRAS.412.2763F} {412, 2763}

\bibitem[\protect\citeauthoryear{{Freire} et~al.,}{{Freire}
  et~al.}{2012}]{fwe+12}
{Freire} P.~C.~C.,  et~al., 2012, \mn@doi [MNRAS]
  {10.1111/j.1365-2966.2012.21253.x}, \href
  {http://adsabs.harvard.edu/abs/2012MNRAS.423.3328F} {423, 3328}

\bibitem[\protect\citeauthoryear{{Gravity Collaboration} et~al.,}{{Gravity
  Collaboration} et~al.}{2018}]{gra+18}
{Gravity Collaboration} et~al., 2018, \mn@doi [\aap]
  {10.1051/0004-6361/201833718}, \href
  {https://ui.adsabs.harvard.edu/abs/2018A&A...615L..15G} {615, L15}

\bibitem[\protect\citeauthoryear{{Holmberg} \& {Flynn}}{{Holmberg} \&
  {Flynn}}{2004}]{hf04}
{Holmberg} J.,  {Flynn} C.,  2004, \mn@doi [\mnras]
  {10.1111/j.1365-2966.2004.07931.x}, \href
  {https://ui.adsabs.harvard.edu/abs/2004MNRAS.352..440H} {352, 440}

\bibitem[\protect\citeauthoryear{{Hotan}, {van Straten}  \&
  {Manchester}}{{Hotan} et~al.}{2004}]{hvm04}
{Hotan} A.~W.,  {van Straten} W.,   {Manchester} R.~N.,  2004, PASA, 21, 302

\bibitem[\protect\citeauthoryear{{Iben} \& {Livio}}{{Iben} \&
  {Livio}}{1993}]{il93}
{Iben} Icko J.,  {Livio} M.,  1993, \mn@doi [\pasp] {10.1086/133321}, \href
  {https://ui.adsabs.harvard.edu/abs/1993PASP..105.1373I} {105, 1373}

\bibitem[\protect\citeauthoryear{{Ivanova} et~al.,}{{Ivanova}
  et~al.}{2013}]{ijc+13}
{Ivanova} N.,  et~al., 2013, \mn@doi [\aapr] {10.1007/s00159-013-0059-2}, \href
  {https://ui.adsabs.harvard.edu/abs/2013A&ARv..21...59I} {21, 59}

\bibitem[\protect\citeauthoryear{{Karakas}, {Lattanzio}  \& {Pols}}{{Karakas}
  et~al.}{2002}]{klp02}
{Karakas} A.~I.,  {Lattanzio} J.~C.,   {Pols} O.~R.,  2002, \mn@doi [\pasa]
  {10.1071/AS02013}, \href
  {https://ui.adsabs.harvard.edu/abs/2002PASA...19..515K} {19, 515}

\bibitem[\protect\citeauthoryear{Kopeikin}{Kopeikin}{1995}]{kop95}
Kopeikin S.~M.,  1995, ApJ, 439, L5

\bibitem[\protect\citeauthoryear{Kopeikin}{Kopeikin}{1996}]{kop96}
Kopeikin S.~M.,  1996, ApJ, 467, L93

\bibitem[\protect\citeauthoryear{{Kramer} et~al.,}{{Kramer}
  et~al.}{2006}]{ksm+06}
{Kramer} M.,  et~al., 2006, \mn@doi [Science] {10.1126/science.1132305}, \href
  {https://ui.adsabs.harvard.edu/abs/2006Sci...314...97K} {314, 97}

\bibitem[\protect\citeauthoryear{{Kramer}, {Stairs}  \& {Venkatraman
  Krishnan}}{{Kramer} et~al.}{2020}]{ksv+20}
{Kramer} M.,  {Stairs} I.,   {Venkatraman Krishnan} V.,  2020, MNRAS

\bibitem[\protect\citeauthoryear{{Lattimer} \& {Prakash}}{{Lattimer} \&
  {Prakash}}{2001}]{lp01}
{Lattimer} J.~M.,  {Prakash} M.,  2001, ApJ, 550, 426

\bibitem[\protect\citeauthoryear{{Lazaridis} et~al.,}{{Lazaridis}
  et~al.}{2009}]{lwj+09}
{Lazaridis} K.,  et~al., 2009, \mn@doi [\mnras]
  {10.1111/j.1365-2966.2009.15481.x}, \href
  {https://ui.adsabs.harvard.edu/abs/2009MNRAS.400..805L} {400, 805}

\bibitem[\protect\citeauthoryear{{Lazarus}, {Karuppusamy}, {Graikou},
  {Caballero}, {Champion}, {Lee}, {Verbiest}  \& {Kramer}}{{Lazarus}
  et~al.}{2016}]{lkg+16}
{Lazarus} P.,  {Karuppusamy} R.,  {Graikou} E.,  {Caballero} R.~N.,  {Champion}
  D.~J.,  {Lee} K.~J.,  {Verbiest} J.~P.~W.,   {Kramer} M.,  2016, \mn@doi
  [MNRAS] {10.1093/mnras/stw189}, \href
  {http://adsabs.harvard.edu/abs/2016MNRAS.458..868L} {458, 868}

\bibitem[\protect\citeauthoryear{{Liu} et~al.,}{{Liu} et~al.}{2014}]{ldc+14}
{Liu} K.,  et~al., 2014, \mn@doi [\mnras] {10.1093/mnras/stu1420}, \href
  {https://ui.adsabs.harvard.edu/abs/2014MNRAS.443.3752L} {443, 3752}

\bibitem[\protect\citeauthoryear{{Lorimer}}{{Lorimer}}{2008}]{lor08}
{Lorimer} D.~R.,  2008, \mn@doi [Living Reviews in Relativity]
  {10.12942/lrr-2008-8}, \href
  {https://ui.adsabs.harvard.edu/abs/2008LRR....11....8L} {11, 8}

\bibitem[\protect\citeauthoryear{Lorimer \& Kramer}{Lorimer \&
  Kramer}{2005}]{lk05}
Lorimer D.~R.,  Kramer M.,  2005, Handbook of Pulsar Astronomy.
Cambridge University Press

\bibitem[\protect\citeauthoryear{{Martinez} et~al.,}{{Martinez}
  et~al.}{2015}]{msf+15}
{Martinez} J.~G.,  et~al., 2015, \mn@doi [\apj] {10.1088/0004-637X/812/2/143},
  \href {https://ui.adsabs.harvard.edu/abs/2015ApJ...812..143M} {812, 143}

\bibitem[\protect\citeauthoryear{{McGaugh}}{{McGaugh}}{2018}]{mcg18}
{McGaugh} S.~S.,  2018, \mn@doi [Research Notes of the American Astronomical
  Society] {10.3847/2515-5172/aadd4b}, \href
  {https://ui.adsabs.harvard.edu/abs/2018RNAAS...2..156M} {2, 156}

\bibitem[\protect\citeauthoryear{{Misra}, {Fragos}, {Tauris}, {Zapartas}  \&
  {Aguilera-Dena}}{{Misra} et~al.}{2020}]{mft+20}
{Misra} D.,  {Fragos} T.,  {Tauris} T.,  {Zapartas} E.,   {Aguilera-Dena}
  D.~R.,  2020, arXiv e-prints, \href
  {https://ui.adsabs.harvard.edu/abs/2020arXiv200401205M} {p. arXiv:2004.01205}

\bibitem[\protect\citeauthoryear{{Ng} et~al.,}{{Ng} et~al.}{2014}]{nbb+14}
{Ng} C.,  et~al., 2014, \mn@doi [\mnras] {10.1093/mnras/stu067}, \href
  {https://ui.adsabs.harvard.edu/abs/2014MNRAS.439.1865N} {439, 1865}

\bibitem[\protect\citeauthoryear{{Ng} et~al.,}{{Ng} et~al.}{2018}]{nct+18}
{Ng} C.,  et~al., 2018, \mn@doi [\mnras] {10.1093/mnras/sty482}, \href
  {https://ui.adsabs.harvard.edu/abs/2018MNRAS.476.4315N} {476, 4315}

\bibitem[\protect\citeauthoryear{{{\"O}zel} \& {Freire}}{{{\"O}zel} \&
  {Freire}}{2016}]{of16}
{{\"O}zel} F.,  {Freire} P.,  2016, \mn@doi [\araa]
  {10.1146/annurev-astro-081915-023322}, \href
  {https://ui.adsabs.harvard.edu/abs/2016ARA&A..54..401O} {54, 401}

\bibitem[\protect\citeauthoryear{{Paczynski}}{{Paczynski}}{1976}]{pac76}
{Paczynski} B.,  1976, in {Eggleton} P.,  {Mitton} S.,   {Whelan} J.,  eds,
  IAU Symposium Vol. 73, Structure and Evolution of Close Binary Systems. p.~75

\bibitem[\protect\citeauthoryear{{Parent} et~al.,}{{Parent}
  et~al.}{2019}]{pkr+19}
{Parent} E.,  et~al., 2019, \mn@doi [\apj] {10.3847/1538-4357/ab4f85}, \href
  {https://ui.adsabs.harvard.edu/abs/2019ApJ...886..148P} {886, 148}

\bibitem[\protect\citeauthoryear{{Peters} \& {Mathews}}{{Peters} \&
  {Mathews}}{1963}]{pm63}
{Peters} P.~C.,  {Mathews} J.,  1963, \mn@doi [Physical Review]
  {10.1103/PhysRev.131.435}, \href
  {https://ui.adsabs.harvard.edu/abs/1963PhRv..131..435P} {131, 435}

\bibitem[\protect\citeauthoryear{Phinney}{Phinney}{1992}]{phi92}
Phinney E.~S.,  1992, 341, 39

\bibitem[\protect\citeauthoryear{Phinney \& Kulkarni}{Phinney \&
  Kulkarni}{1994}]{pk94}
Phinney E.~S.,  Kulkarni S.~R.,  1994, ARA\&A, 32, 591

\bibitem[\protect\citeauthoryear{{Radhakrishnan} \& {Cooke}}{{Radhakrishnan} \&
  {Cooke}}{1969}]{rc69}
{Radhakrishnan} V.,  {Cooke} D.~J.,  1969, Ap. Lett., 3, 225

\bibitem[\protect\citeauthoryear{{Shaifullah} et~al.,}{{Shaifullah}
  et~al.}{2016}]{svf+16}
{Shaifullah} G.,  et~al., 2016, \mn@doi [\mnras] {10.1093/mnras/stw1737}, \href
  {https://ui.adsabs.harvard.edu/abs/2016MNRAS.462.1029S} {462, 1029}

\bibitem[\protect\citeauthoryear{{Shao}, {Sennett}, {Buonanno}, {Kramer}  \&
  {Wex}}{{Shao} et~al.}{2017}]{ssb+17}
{Shao} L.,  {Sennett} N.,  {Buonanno} A.,  {Kramer} M.,   {Wex} N.,  2017,
  \mn@doi [Physical Review X] {10.1103/PhysRevX.7.041025}, \href
  {https://ui.adsabs.harvard.edu/abs/2017PhRvX...7d1025S} {7, 041025}

\bibitem[\protect\citeauthoryear{Shapiro}{Shapiro}{1964}]{sha64}
Shapiro I.~I.,  1964, PRL, 13, 789

\bibitem[\protect\citeauthoryear{Shklovskii}{Shklovskii}{1970}]{shk70}
Shklovskii I.~S.,  1970, 13, 562

\bibitem[\protect\citeauthoryear{{Stairs}}{{Stairs}}{2003}]{sta03}
{Stairs} I.~H.,  2003, \mn@doi [Living Reviews in Relativity]
  {10.12942/lrr-2003-5}, \href
  {https://ui.adsabs.harvard.edu/abs/2003LRR.....6....5S} {6, 5}

\bibitem[\protect\citeauthoryear{{Stovall} et~al.,}{{Stovall}
  et~al.}{2019}]{sfa+19}
{Stovall} K.,  et~al., 2019, \mn@doi [\apj] {10.3847/1538-4357/aaf37d}, \href
  {https://ui.adsabs.harvard.edu/abs/2019ApJ...870...74S} {870, 74}

\bibitem[\protect\citeauthoryear{{Taam}, {King}  \& {Ritter}}{{Taam}
  et~al.}{2000}]{tkr00}
{Taam} R.~E.,  {King} A.~R.,   {Ritter} H.,  2000, ApJ, 541, 329

\bibitem[\protect\citeauthoryear{{Tauris} \& {Janka}}{{Tauris} \&
  {Janka}}{2019}]{tj19}
{Tauris} T.~M.,  {Janka} H.-T.,  2019, \mn@doi [\apjl]
  {10.3847/2041-8213/ab5642}, \href
  {https://ui.adsabs.harvard.edu/abs/2019ApJ...886L..20T} {886, L20}

\bibitem[\protect\citeauthoryear{{Tauris} \& {Savonije}}{{Tauris} \&
  {Savonije}}{1999}]{ts99}
{Tauris} T.~M.,  {Savonije} G.~J.,  1999, \aap, \href
  {https://ui.adsabs.harvard.edu/abs/1999A&A...350..928T} {350, 928}

\bibitem[\protect\citeauthoryear{{Tauris}, {Langer}  \& {Kramer}}{{Tauris}
  et~al.}{2011}]{tlk11}
{Tauris} T.~M.,  {Langer} N.,   {Kramer} M.,  2011, \mn@doi [\mnras]
  {10.1111/j.1365-2966.2011.19189.x}, \href
  {https://ui.adsabs.harvard.edu/abs/2011MNRAS.416.2130T} {416, 2130}

\bibitem[\protect\citeauthoryear{{Tauris}, {Langer}  \& {Kramer}}{{Tauris}
  et~al.}{2012}]{tlk12}
{Tauris} T.~M.,  {Langer} N.,   {Kramer} M.,  2012, \mn@doi [\mnras]
  {10.1111/j.1365-2966.2012.21446.x}, \href
  {https://ui.adsabs.harvard.edu/abs/2012MNRAS.425.1601T} {425, 1601}

\bibitem[\protect\citeauthoryear{{Tauris}, {Langer}  \&
  {Podsiadlowski}}{{Tauris} et~al.}{2015}]{tlp+15}
{Tauris} T.~M.,  {Langer} N.,   {Podsiadlowski} P.,  2015, \mn@doi [\mnras]
  {10.1093/mnras/stv990}, \href
  {https://ui.adsabs.harvard.edu/abs/2015MNRAS.451.2123T} {451, 2123}

\bibitem[\protect\citeauthoryear{{Tauris} et~al.,}{{Tauris}
  et~al.}{2017}]{tkf+17}
{Tauris} T.~M.,  et~al., 2017, \mn@doi [\apj] {10.3847/1538-4357/aa7e89}, \href
  {https://ui.adsabs.harvard.edu/abs/2017ApJ...846..170T} {846, 170}

\bibitem[\protect\citeauthoryear{{Taylor} \& {Weisberg}}{{Taylor} \&
  {Weisberg}}{1982}]{tw82}
{Taylor} J.~H.,  {Weisberg} J.~M.,  1982, \mn@doi [\apj] {10.1086/159690},
  \href {https://ui.adsabs.harvard.edu/abs/1982ApJ...253..908T} {253, 908}

\bibitem[\protect\citeauthoryear{{Venkatraman Krishnan} et~al.,}{{Venkatraman
  Krishnan} et~al.}{2020}]{vbs+20}
{Venkatraman Krishnan} V.,  et~al., 2020, \mn@doi [Science]
  {10.1126/science.aax7007}, \href
  {https://ui.adsabs.harvard.edu/abs/2020Sci...367..577V} {367, 577}

\bibitem[\protect\citeauthoryear{{Voisin}, {Cognard}, {Freire}, {Wex},
  {Guillemot}, {Desvignes}, {Kramer}  \& {Theureau}}{{Voisin}
  et~al.}{2020}]{vfc+20}
{Voisin} G.,  {Cognard} I.,  {Freire} P.~C.~C.,  {Wex} N.,  {Guillemot} L.,
  {Desvignes} G.,  {Kramer} M.,   {Theureau} G.,  2020, \mn@doi [\aap]
  {10.1051/0004-6361/202038104}, \href
  {https://ui.adsabs.harvard.edu/abs/2020A&A...638A..24V} {638, A24}

\bibitem[\protect\citeauthoryear{{Weisberg} \& {Huang}}{{Weisberg} \&
  {Huang}}{2016}]{wh16}
{Weisberg} J.~M.,  {Huang} Y.,  2016, \mn@doi [\apj]
  {10.3847/0004-637X/829/1/55}, \href
  {https://ui.adsabs.harvard.edu/abs/2016ApJ...829...55W} {829, 55}

\bibitem[\protect\citeauthoryear{Weisberg \& Taylor}{Weisberg \&
  Taylor}{1981}]{wt81}
Weisberg J.,  Taylor J.,  1981, Gen. Relativ. Gravit., 13, 1

\bibitem[\protect\citeauthoryear{{Wex}}{{Wex}}{2014}]{wex14}
{Wex} N.,  2014, arXiv e-prints, \href
  {https://ui.adsabs.harvard.edu/abs/2014arXiv1402.5594W} {p. arXiv:1402.5594}

\bibitem[\protect\citeauthoryear{{Yao}, {Manchester}  \& {Wang}}{{Yao}
  et~al.}{2017}]{ymw17}
{Yao} J.~M.,  {Manchester} R.~N.,   {Wang} N.,  2017, \mn@doi [\apj]
  {10.3847/1538-4357/835/1/29}, \href
  {https://ui.adsabs.harvard.edu/abs/2017ApJ...835...29Y} {835, 29}

\bibitem[\protect\citeauthoryear{{Zhu} et~al.,}{{Zhu} et~al.}{2019}]{zfk+19}
{Zhu} W.~W.,  et~al., 2019, \mn@doi [\apj] {10.3847/1538-4357/ab2bef}, \href
  {https://ui.adsabs.harvard.edu/abs/2019ApJ...881..165Z} {881, 165}

\bibitem[\protect\citeauthoryear{{de Kool}}{{de Kool}}{1990}]{dek90}
{de Kool} M.,  1990, \mn@doi [\apj] {10.1086/168974}, \href
  {https://ui.adsabs.harvard.edu/abs/1990ApJ...358..189D} {358, 189}

\bibitem[\protect\citeauthoryear{{van der Sluys}, {Verbunt}  \& {Pols}}{{van
  der Sluys} et~al.}{2006}]{vvp06}
{van der Sluys} M.~V.,  {Verbunt} F.,   {Pols} O.~R.,  2006, \mn@doi [\aap]
  {10.1051/0004-6361:20065066}, \href
  {https://ui.adsabs.harvard.edu/abs/2006A&A...460..209V} {460, 209}

\makeatother
\end{thebibliography}
\bsp	

\label{lastpage}
\end{document}